\documentclass[useAMS,usenatbib]{mn2e}

\newcommand{\Msun}{\mbox{${\rm M}_\odot$}}
\newcommand{\hMsun}{\mbox{$h^{-1} {\rm M}_\odot$}}
\newcommand{\bfMsun}{\mbox{$\mathbf{{\rm \bf M}_\odot}$}}
\newcommand{\LCDM}{\mbox{$\Lambda$CDM }}
\def\apgt{\ {\raise-.5ex\hbox{$\buildrel>\over\sim$}}\ }
\def\hkpc{~h^{-1} {\rm kpc}}
\def\Mpch{~h^{-1} {\rm Mpc}}

\usepackage[utf8]{inputenc}
\usepackage{graphicx}
\usepackage{subfig}

\title{Evolution of star clusters in a cosmological tidal field}
\author[Rieder et al.]
{Steven~Rieder$^{1,2}$\thanks{rieder@strw.leidenuniv.nl},
 Tomoaki~Ishiyama$^{3}$,
 Paul~Langelaan$^{1}$,
 Junichiro~Makino$^{4}$,
 \newauthor
 Stephen~L.~W.~McMillan$^{5}$
 and Simon~Portegies~Zwart$^{1}$\\
\\
  $^1$ Sterrewacht Leiden, Leiden University, P.O. Box 9513, 2300 RA Leiden, The Netherlands\\
  $^2$ Section System and Network Engineering, University of Amsterdam, Amsterdam, The Netherlands\\
  $^3$ Center for Computational Sciences, University of Tsukuba, Japan\\
  $^4$ Interactive Research Center of Science, Graduate School of Science and Engineering Tokyo Institute of Technology,\\
  2-12-1 Ookayama, Meguro, Tokyo 152-8551, Japan\\
  $^5$ Department of Physics, Drexel University, Philadelphia, PA 19104, USA}

\date{Accepted .... Received ...; in original form ...}

\pagerange{\pageref{firstpage}--\pageref{lastpage}} \pubyear{}

\begin{document}

\label{firstpage}

\maketitle

\begin{abstract}

  We present a method to couple $N$-body star cluster simulations to a
  cosmological tidal field, using the Astrophysical Multipurpose
  Software Environment. We apply this method to star clusters embedded
  in the CosmoGrid dark matter-only \LCDM\,simulation.  Our star
  clusters are born at $z=10$ (corresponding to an age of the Universe
  of about 500\,Myr) by selecting a dark matter particle and
  initializing a star cluster with 32,000 stars on its location. We
  then follow the dynamical evolution of the star cluster within the
  cosmological environment.

  We compare the evolution of star clusters in two Milky-Way size
  haloes with a different accretion history.  The mass loss of the
  star clusters is continuous irrespective of the tidal history of the
  host halo, but major merger events tend to increase the rate of mass
  loss. From the selected two dark matter haloes, the halo that
  experienced the larger number of mergers tends to drive a smaller
  mass loss rate from the embedded star clusters, even though the
  final masses of both haloes are similar.  
  We identify two families of star clusters: native clusters, which
  become part of the main halo before its final major merger event,
  and the immigrant clusters, which are accreted upon or after this
  event; native clusters tend to evaporate more quickly than immigrant
  clusters. Accounting for the evolution of the dark matter halo
  causes immigrant star clusters to retain more mass than when the z=0
  tidal field is taken as a static potential.  The reason for this is
  the weaker tidal field experienced by immigrant star clusters before
  merging with the larger dark matter halo.

\end{abstract}


\section{Introduction}

Globular clusters are dense self gravitating systems of a few $10^4$
to $\sim 10^6$ stars \citep{2006ARA&A..44..193B}.  With an age of
about 12.6\,Gyr \citep{2003Sci...299...65K} they are among the oldest
objects in the universe and tend to populate the haloes of galaxies.
Their age is consistent with being born between $z=12$ to 7, which is
consistent with the results of \LCDM simulations
\citep{2005ApJ...623..650K}.

From the time the clusters were born on galaxies grow by about a
factor 100 in mass via mergers to their current mass, size and
morphology \citep{1978MNRAS.183..341W,1999coph.book.....P}.  The
environment in which the globular clusters evolved since their birth
has consequently changed quite dramatically over their lifetimes.
These changes may have a profound effect on the evolution of star
clusters.

Most modern star-cluster simulations take some sort of background
potential of the host galaxy into account. This started already in the
early 1990s with \cite{1990ApJ...351..121C}, and soon afterwards
became a lively industry. Many simulations have been performed with a
fixed tidal limit \citep{1997MNRAS.289..898V, 1998A&A...337..363P,
2001MNRAS.324..218G} whereas other include some sort of tidal
potential with a more fluent description of the tidal field
\citep{1997ApJ...474..223G, 2000ApJ...535..759T, 2010MNRAS.409..305L}.
In most of these simulations the cluster orbit was circular and did
not change with time. In a few cases the orbit was allowed to be
eccentric, but still did not change with time
\citep{2003MNRAS.340..227B, 2011MNRAS.410.2698G, 2009MNRAS.395.1173G}.
The next refinement was the relaxing of the orbital parameters,
allowing the cluster orbit to change \citep{2005PASJ...57..155T,
2010PASJ...62.1215T, 2010NewA...15...46P, 2011MNRAS.418..759R,
2012MNRAS.419.3244B, 2013MNRAS.431L..83R}.  The evolution of star
clusters in a live galactic potential combined with parametrized
cluster evolution was studied by \cite{2011MNRAS.414.1339K}, and
\cite{2012ApJ...746...26M} performed simulations in which they resolve
the formation of star clusters in a single galaxy merger event.

However, the mass evolution of the parent galaxy is generally ignored
in all these simulations, except for the few cases with a galaxy
merger \citep{2009PASJ...61..481S, 2011MNRAS.418..759R,
2012MNRAS.421.1927K, 2013MNRAS.431L..83R}.

In this paper we study the evolution of star clusters in a
cosmological environment. The background potential against which the
star clusters are evolved, are taken from the CosmoGrid dark-matter
only \LCDM\, simulation \citep{2010IEEEC..43...63P}.  We selected two
Milky Way like haloes in which the star clusters are simulated.  The
coupling between the cosmological simulation and the star cluster is
realized via the Astrophysical Multipurpose Software Environment
(AMUSE) \citep{2009NewA...14..369P, 2011arXiv1110.2785P,
2012arXiv1204.5522P, 2013arXiv1307.3016P}.  In these simulations we
initialize a total of 30 star cluster at $z=10$, and evolve them
together with the cosmological simulation up to $z=0$. 

\section{The experimental setup}

We simulate star clusters in a \LCDM environment. We do this in two
steps, first by calculating the \LCDM environment and then using the
tidal field from this environment as an external tidal field for the
star cluster simulations. We investigate the results of two distinct
regions in the \LCDM environment.

\subsection{The CosmoGrid $N$-body simulation}

The cosmological simulation employed in this article originates from
the CosmoGrid calculation \citep{2010IEEEC..43...63P,
2013ApJ...767..146I}, which is a dark matter-only \LCDM simulation of
$2048^3$ particles in a $(21 \Mpch)^3$ co-moving cosmological volume.
We performed these simulations using the {\tt GreeM}
\citep{2009PASJ...61.1319I, 2012arXiv1211.4406I} and {\tt Sushi}
\citep{2011CS&D....4a5001G} codes. GreeM is a massively parallel
TreePM code based on the implementation of \cite{2005PASJ...57..849Y}.
The SUSHI code is an extension of the GreeM code, which can run on a
planet wide grid of supercomputers. Within both codes, the equations
of motion are integrated in co-moving coordinates using the leap-frog
scheme with a shared, adaptive timestep. In this simulation each
particle has a mass of $1.28\times10^5$\,\Msun. In total, we have 556
snapshots, separated by $\mbox{dt} \simeq 35 \mbox{Myr}$ (for $t <
7.5\mbox{Gyr}$) and $\mbox{dt} \simeq 17.5 \mbox{Myr}$ (for $t \geq
7.5\mbox{Gyr}$).  The CosmoGrid simulation lasted from $z=65$ to
$z=0$. We employed the following cosmological parameters: $\Omega_0 =
0.3, \lambda_0 = 0.7, h=0.7, \sigma_8 = 0.8, n = 1.0$. For more
details on the simulation see \cite{2010IEEEC..43...63P,
2013ApJ...767..146I}.

\subsection{Halo catalogue}

We use the halo finder {\tt Rockstar} \citep{2013ApJ...762..109B}
to identify haloes in each snapshot. Rockstar is based on adaptive
hierarchical refinement of friends-of-friends groups in six dimensions
and allows for the robust tracking of subhaloes. We use the
gravitationally consistent merger tree code from
\cite{2013ApJ...763...18B} to construct the merger history for all
haloes identified by Rockstar.  We use AMUSE to find
the radial density profiles for our haloes.

From the $z=0$ haloes, we then select two haloes, based on their
relative isolation and a mass comparable to that of the Milky-Way
Galaxy. The two haloes are quite similar in many respects, but have a
different merger history and the number of subhaloes at $z=0$ is
different.

\begin{figure*}
  \includegraphics[height=0.9\textheight]{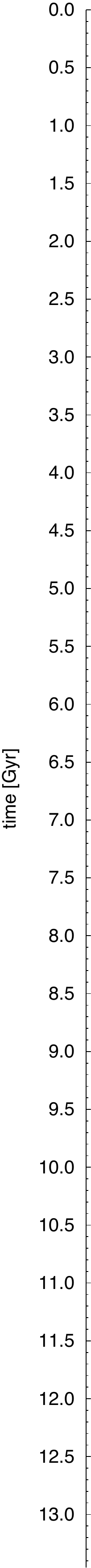}
  \includegraphics[height=0.9\textheight]{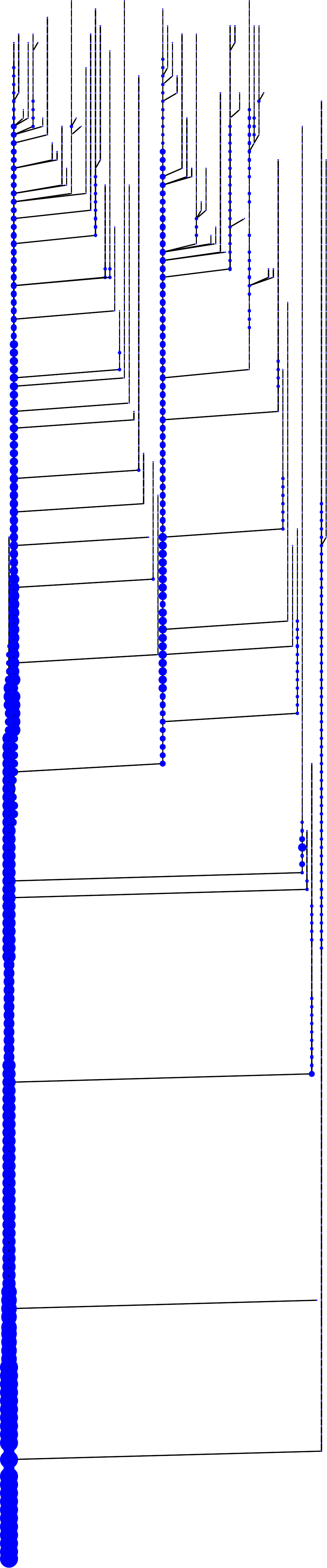}
  \includegraphics[height=0.9\textheight]{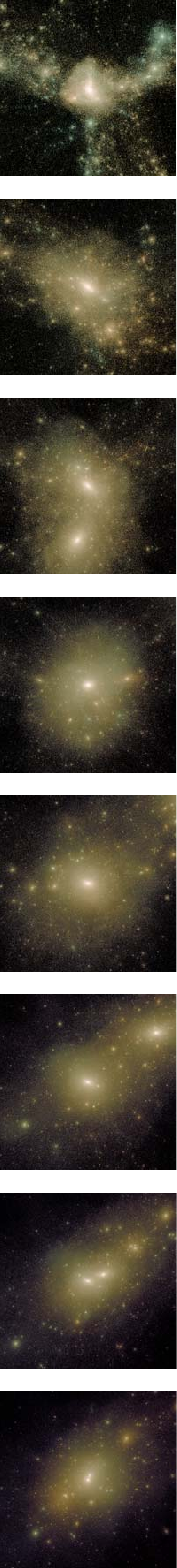}
  \includegraphics[height=0.9\textheight]{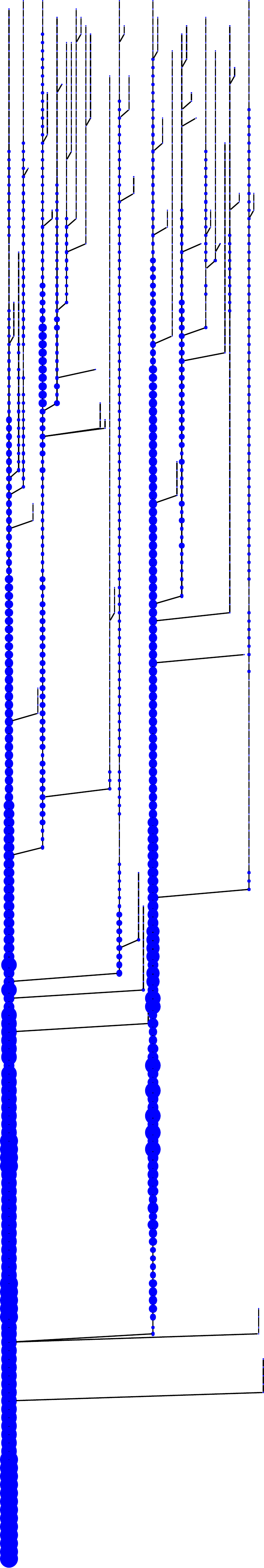}
  \includegraphics[height=0.9\textheight]{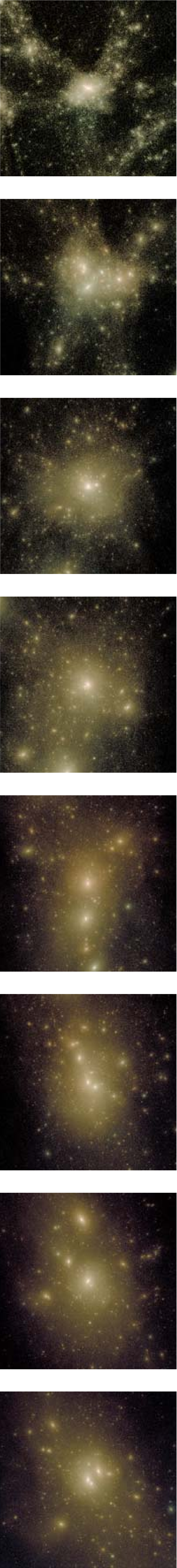}
  \caption[]{The merger history of the two selected dark-matter haloes
  A (left) and B (right). The size of each circle is proportional to
  the virial mass of the halo. Only haloes and subhaloes with a peak
  mass larger than $5\times 10^{8}$\hMsun that are accreted at $z=0$
  are plotted. The bottom two halo images are identical to those in
  Fig.\,\ref{Fig:HaloProjection}.} \label{Fig:MergerHistory}
\end{figure*}

\begin{figure}
  \subfloat[]{
    \includegraphics[width=\columnwidth]{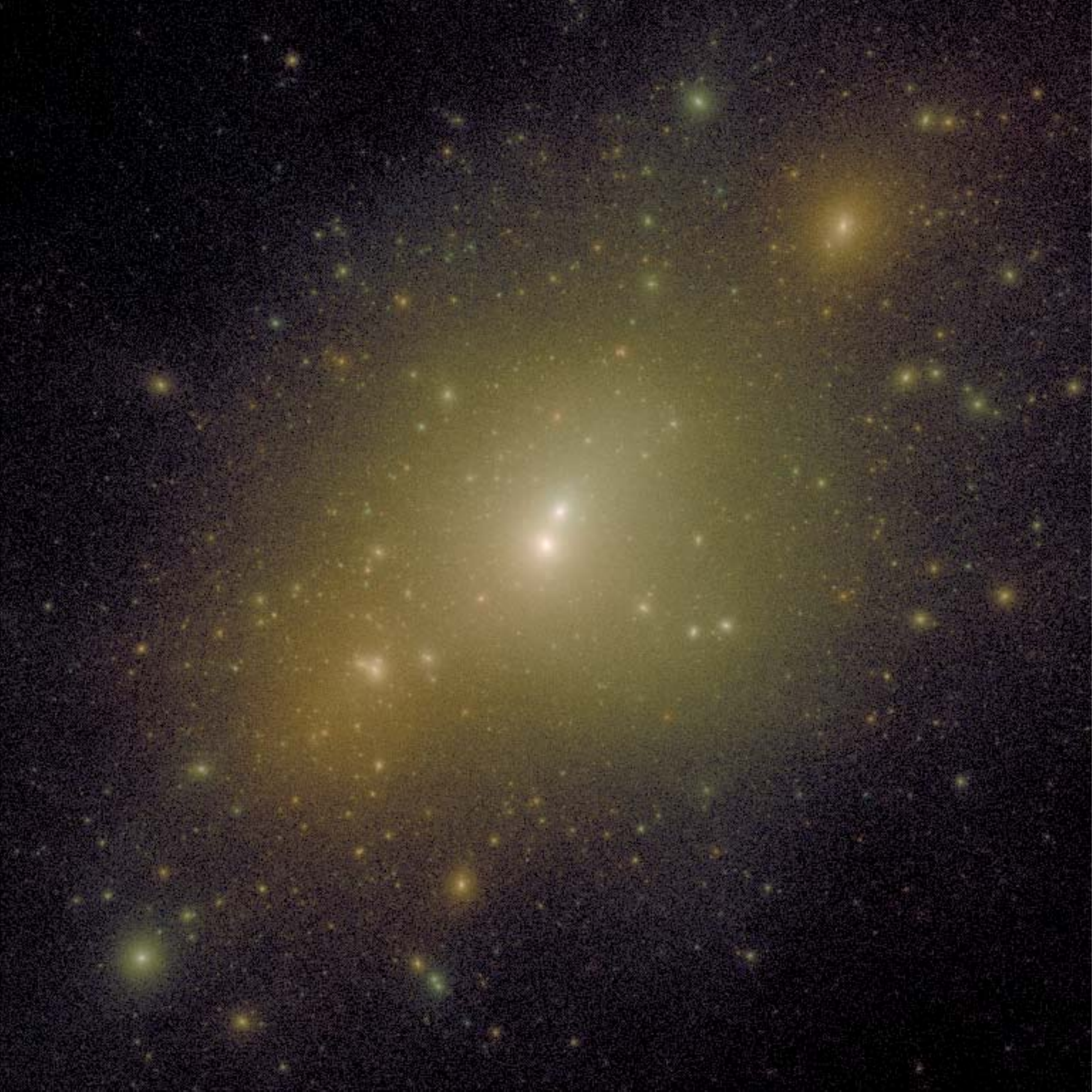}}\\
  \subfloat[]{
    \includegraphics[width=\columnwidth]{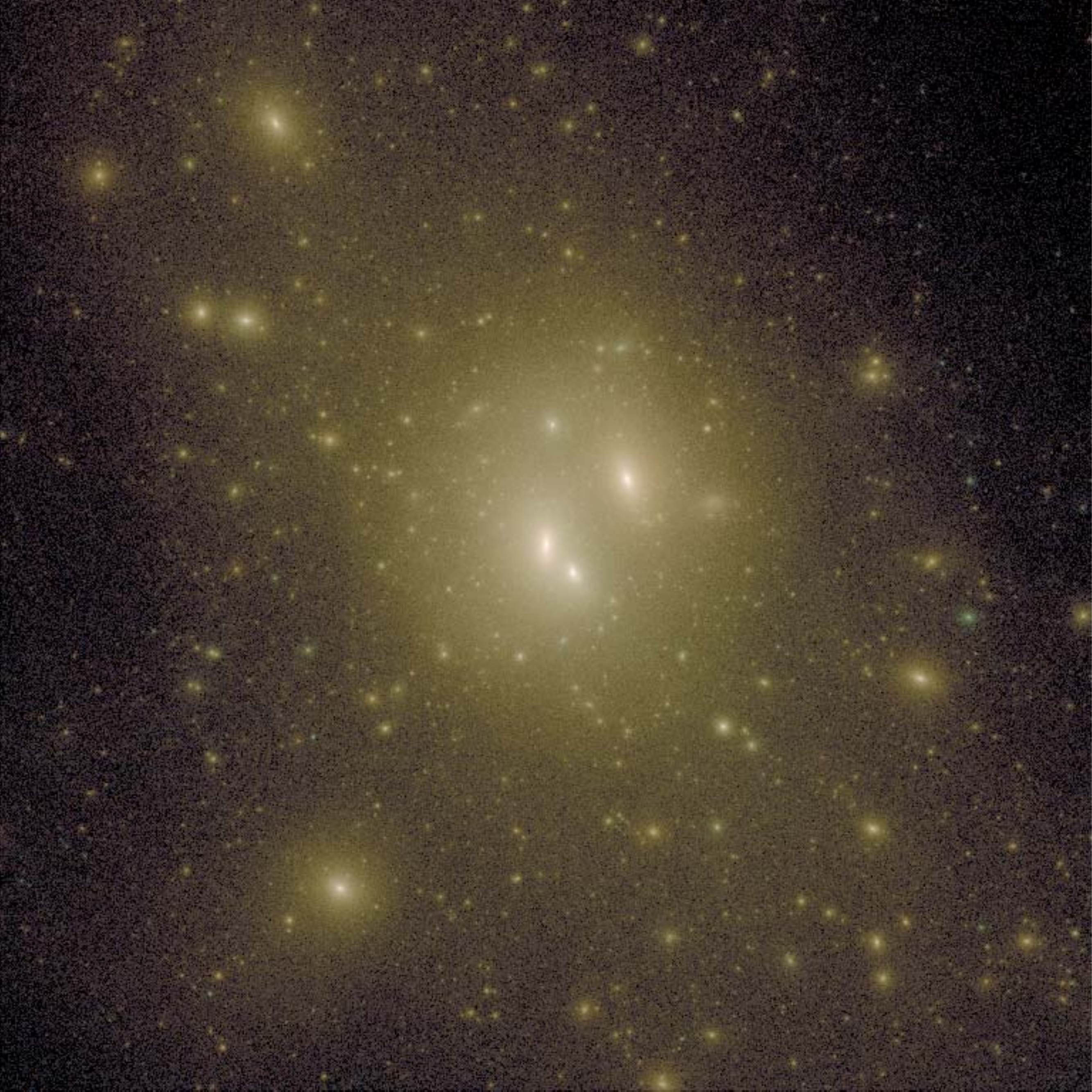}}
  \caption[]{Projected image of the two selected haloes A (top panel)
  and B (bottom panel), in the final snapshot ($z=0$). The linear
  dimension of the image is $400\hkpc$.  The intensity represents the
  column density (scaled to minimum/maximum values) and the colour is
  scaled to the velocity dispersion. Each galaxy halo contains about 5
  million dark-matter particles. }
  \label{Fig:HaloProjection}
\end{figure}

In Fig.\,\ref{Fig:MergerHistory},
we present the merger history of the two haloes, both schematically
and visually. Halo A completes a major merger at around $t=6.5$\,Gyr,
the halo it merges with can be seen in the third figure from the top.
At the end of the simulation (see Fig.\, \ref{Fig:HaloProjection}
a), it is in the process of
merging with another similar-sized halo. Other than that, there is no
significant interaction between $t=6.5$\,Gyr and $t=13.7$\,Gyr. By the
end of the simulation, halo A is the largest object within a radius of
$6.3\Mpch$ (see Tab.\,\ref{Tab:HaloProperties}).  Halo B exists 
in a denser part of the volume, and as a result more structure is
visible in the outskirts of its environment (see
Fig.\,\ref{Fig:HaloProjection}b). By $z=0$, it is the largest 
halo within only $0.97\Mpch$. During its history, it underwent many
small merger events, and one long-lasting major merger event that
completed around $t=11$\,Gyr. 

\begin{figure}
  \includegraphics[width=\columnwidth]{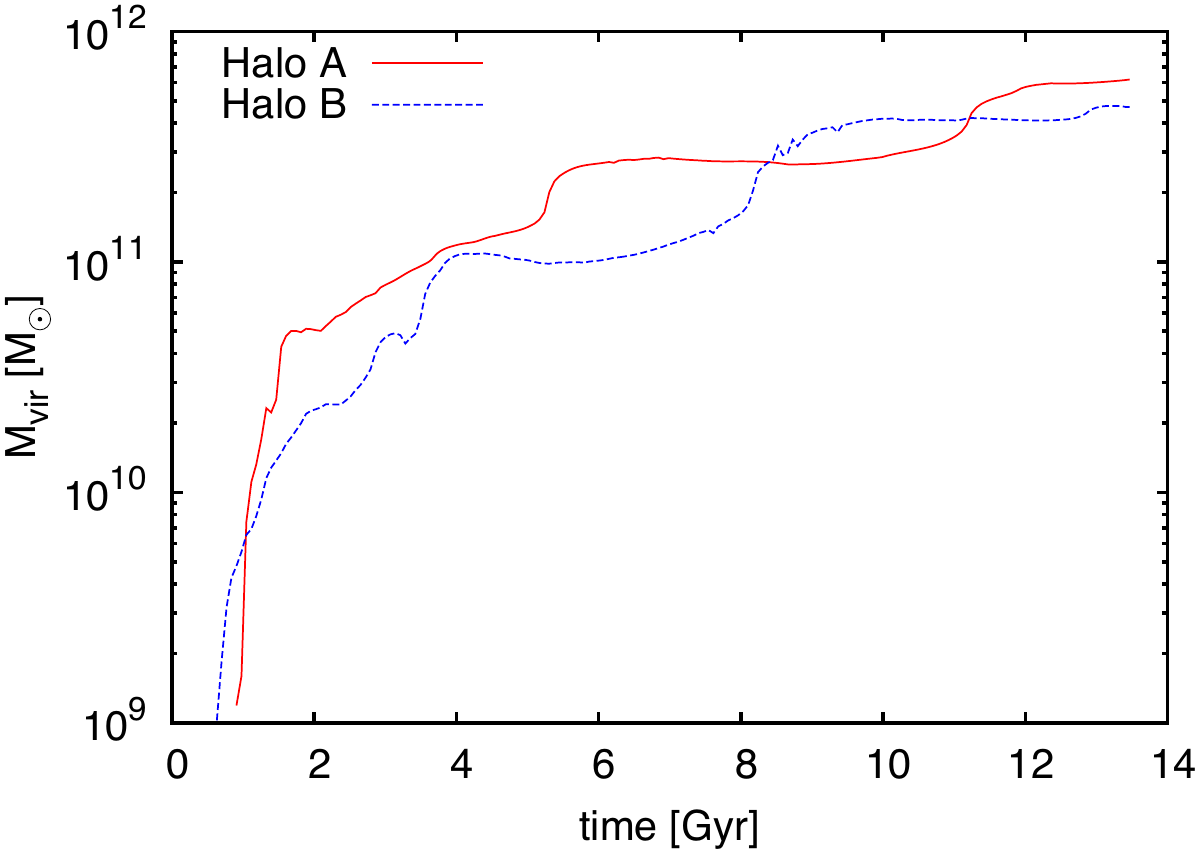}
  \caption[]{Evolution of the virial mass of the two selected haloes A
  and B. Halo A experiences major merger events around 5\,Gyr and
  11\,Gyr, while halo B a long-lasting major merger event from around
  8\,Gyr on. } \label{Fig:HaloMassEvolution}
\end{figure}

In the mass evolution of both haloes
(Fig.\,\ref{Fig:HaloMassEvolution}), the larger merger events are
clearly visible. Since the virial mass includes the mass from
subhaloes, the mergers are visible here at the start of interaction,
rather than at the end as in Fig.\,\ref{Fig:MergerHistory}.

\begin{figure}
  \includegraphics[width=\columnwidth]{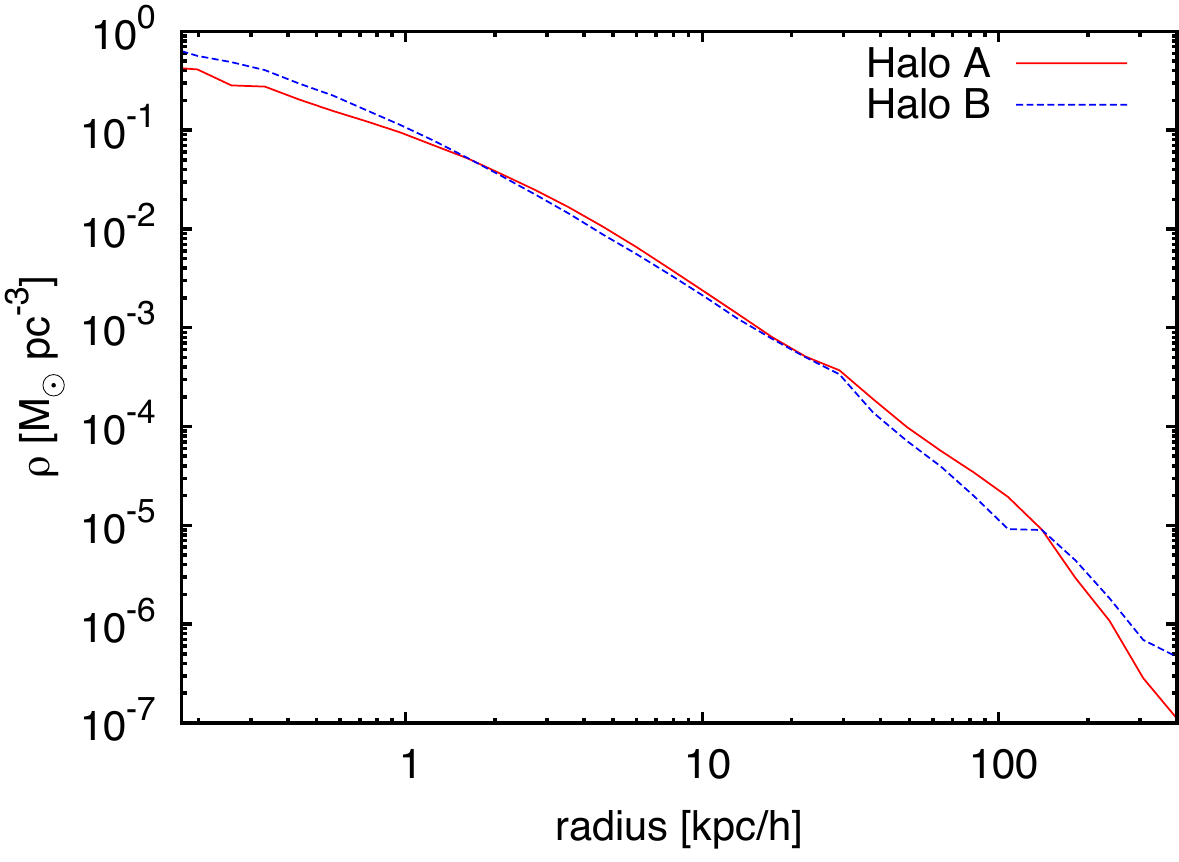}
  \caption[]{Radial density profiles of the two selected haloes A and
  B and their environment out to $400\hkpc$ at $z=0$. The clusters are
  selected at galactocentric radii of $3$, $6$ and $12\hkpc$.}
\label{Fig:RadialProfiles}
\end{figure}

The halo density profiles at $z=0$ (see
Fig.\,\ref{Fig:RadialProfiles}) are consistent with the haloes
described in \cite{2013ApJ...767..146I}, with concentration parameters
$c_{\rm vmax}$ of $3.53$ and $3.76$ for haloes A and B respectively.
The lower concentration of halo A may be explained by its ongoing
major merger event, which causes the halo to have two cores (see
Fig.\,\ref{Fig:HaloProjection}a).

\begin{table*}
  \begin{minipage}{115mm}
  \caption{Properties of the selected haloes in the final CosmoGrid
  snapshot.}
  \label{Tab:HaloProperties}
  \begin{tabular}{l|l|l|l|l|l|l|l|l}
    \hline \hline
    Halo & $M_{\rm vir}$    & $R_{\rm vir}$     & $V_{\rm max}$ & $c_{\rm vmax}$ & $N_{\rm sub}$ & $D_{\rm n}$       & b/a  & c/a\\
         & $10^{11}$ \hMsun & $h^{-1} {\rm kpc}$& km/s          &                &               & $h^{-1} {\rm Mpc}$&      &    \\
    \hline
    A    & $6.33$           & $173.8$           & 140.2         & 3.53           & 61            & 6.30              & 0.81 & 0.63\\
    B    & $4.78$           & $159.4$           & 133.1         & 3.76           & 29            & 0.97              & 0.73 & 0.68\\
    \hline
  \end{tabular}
  
  \medskip
  $M_{\rm vir}$ and $R_{\rm vir}$ are the virial mass and radius
  \citep{1998ApJ...495...80B}, $V_{\rm max}$ is the maximum of the
  rotation curve, $c_{\rm vmax}$ is the concentration parameter,
  $N_{\rm sub}$ is the number of subhaloes with a mass larger than
  $10^8$\,\hMsun, $D_{\rm n}$ is the distance of the nearest more
  massive halo. a, b and c are the principal axes of the halo.
  \end{minipage}
\end{table*}

\subsection{The clusters}

\begin{figure}
  \subfloat[]{
    \includegraphics[width=0.8\columnwidth]{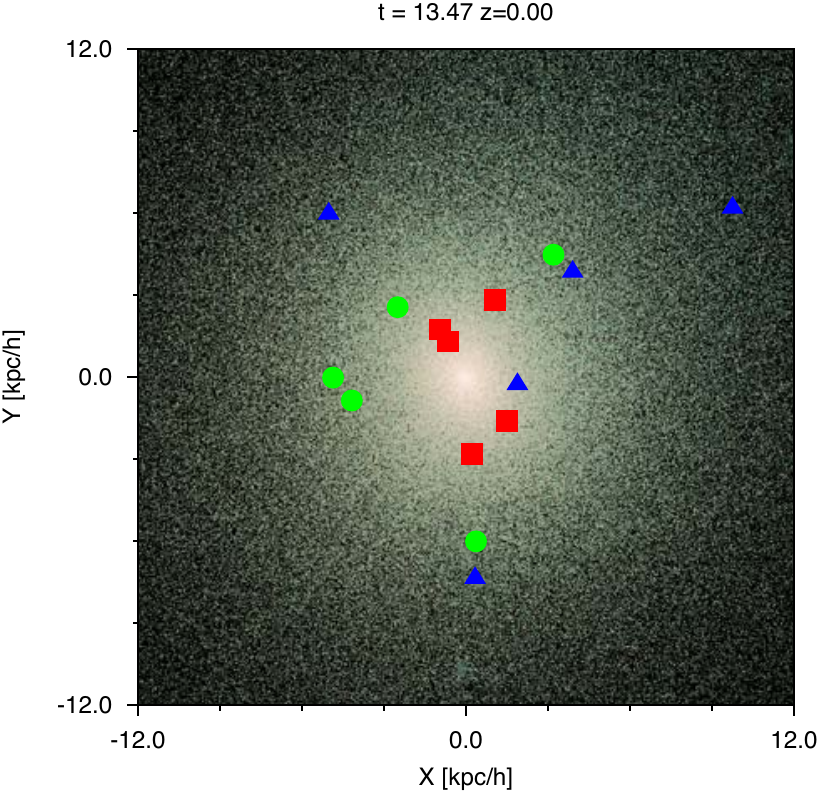}}\\
  \subfloat[]{
    \includegraphics[width=0.8\columnwidth]{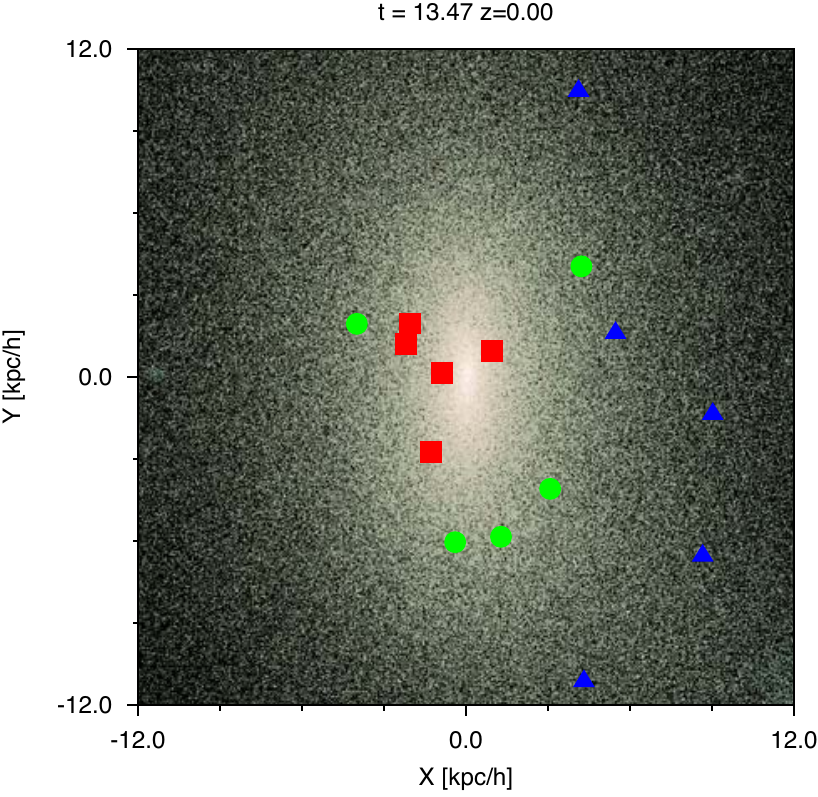}}
  \caption[]{Projected image of the central portion of the two
  selected haloes A (top panel) and B (bottom panel) with their
  `globular clusters' at $z=0$. The image size is $24\hkpc$. Red
  squares indicate clusters at $3\hkpc$ from the halo centre, green
  circles those at $6\hkpc$ and blue triangles those at $12\hkpc$.}
  \label{Fig:HaloZ0Globulars}
\end{figure}

We select a total of 30 dark-matter particles from the $z=0$ snapshot. 
For both haloes we select 15 particles at random, equally
divided over three bins at galactocentric radius $3\pm0.05\hkpc$,
$6\pm0.05\hkpc$ and $12\pm0.05\hkpc$. These selected particles are
considered the globular clusters for which we will calculate the
evolution. Since we do not apply further restrictions in the selection
criteria, the clusters may (and likely will) have their peri- and
apocentres well outside these bins. In Fig.~\ref{Fig:HaloZ0Globulars},
we show a projection of these particles in the central region of their
host halo at $z=0$.

\begin{figure}
  \subfloat[]{
    \includegraphics[width=\columnwidth]{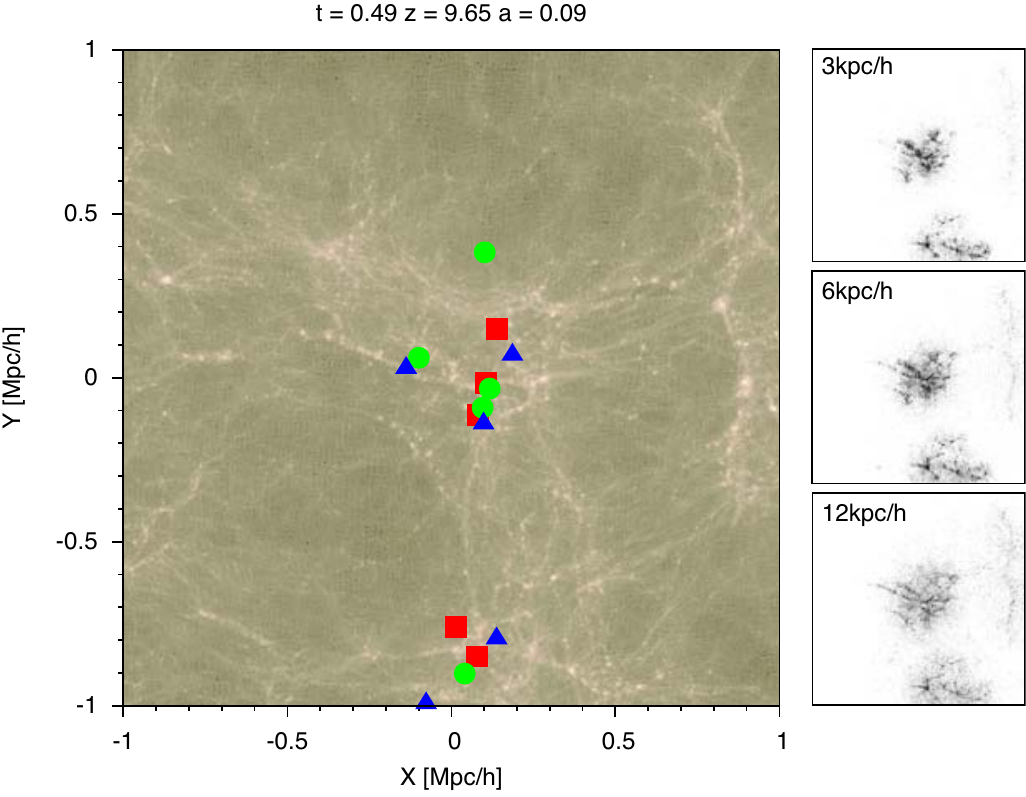}}\\
  \subfloat[]{
    \includegraphics[width=\columnwidth]{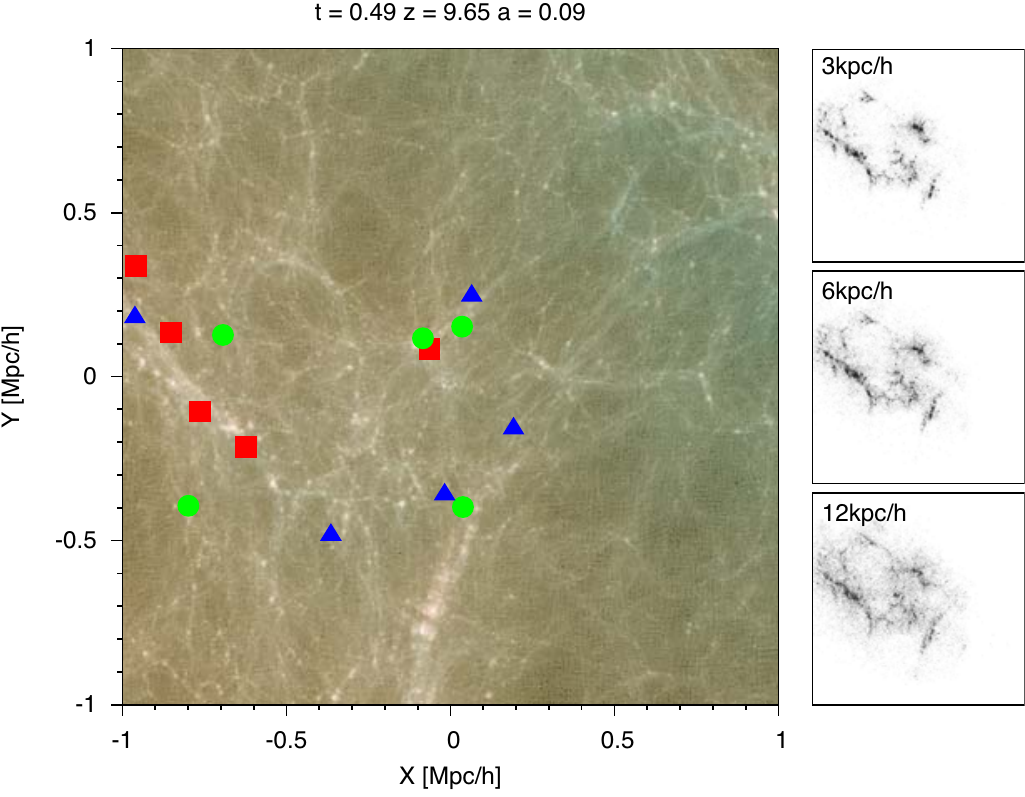}}
  \caption[]{Projected image of halo A (top) and B (bottom) at
    $z=9.65$. The image size is $2\Mpch$. Red squares indicate
    clusters that at $z=0$ were selected at $3\hkpc$ from the halo
    centre, green circles those at $6\hkpc$ and blue triangles those
    at $12\hkpc$. The panels on the right display (from top to bottom)
    the distributions of particles that end up in spherical shells at
    $3$, $6$ and $12 \hkpc$ from the halo centres at $z=0$.}
    \label{Fig:HaloZ10Globulars}
\end{figure}

We locate the selected dark-matter particles at $z=10$, which
corresponds to an age of the universe of $\sim 0.5$\,Gyr.  In
Fig.\,\ref{Fig:HaloZ10Globulars} we present the $z=10$ image of the
two selected haloes with their selected dark-matter particles that
will represent globular clusters, and the distributions from which
these particles are drawn. In both haloes, the particles that end up
in the more central parts of the halo are already the most
concentrated in density centres at $z=10$.  The particles of halo A
are largely concentrated in two regions, in which the haloes that
merge around $t=6.5$\,Gyr form. In contrast, the particles of halo B
are more spread out, reflecting the more violent history of this halo.

At the $z=10$ snapshot we initialize a `globular cluster'. Our
clusters have 32\,000 stars distributed in a
\cite{1911MNRAS..71..460P} sphere with a virial radius of 3\,pc.  The
clusters are assumed to be born in virial equilibrium.  All stars in
the clusters have the same mass and we did not include stellar
evolution.

The star clusters are simulated using the AMUSE framework
\citep{2009NewA...14..369P, 2011arXiv1110.2785P, 2012arXiv1204.5522P,
2013arXiv1307.3016P}.  Our simulation code solves for the equations of
motion using {\tt Bonsai} \citep{2012JCoPh.231.2825B} and {\tt ph4}
\citep{2011arXiv1111.3987M}. Bonsai is a \cite{1986Natur.324..446B}
tree code that runs on GPUs. It supports both shared timesteps and block
timesteps, the latter allowing individual blocks to have different
timesteps for increased accuracy in dense regions without slowing the
simulation down too much. For this code, we adopted an opening angle
(which controls the accuracy, smaller angles being more accurate) of
0.6, we set the smallest timestep to be used to 1/65536 $N$-body time
units \citep{1986LNP...267..233H} and we used a softening length of
0.00125 $N$-body length units (0.00375 parsec).  Ph4 is a direct
$N$-body integrator with block time steps and GPU acceleration. In
order to directly compare the results to those obtained with Bonsai,
we apply the same softening length as before to the runs with ph4.

\begin{figure}
  \includegraphics[width=\columnwidth]{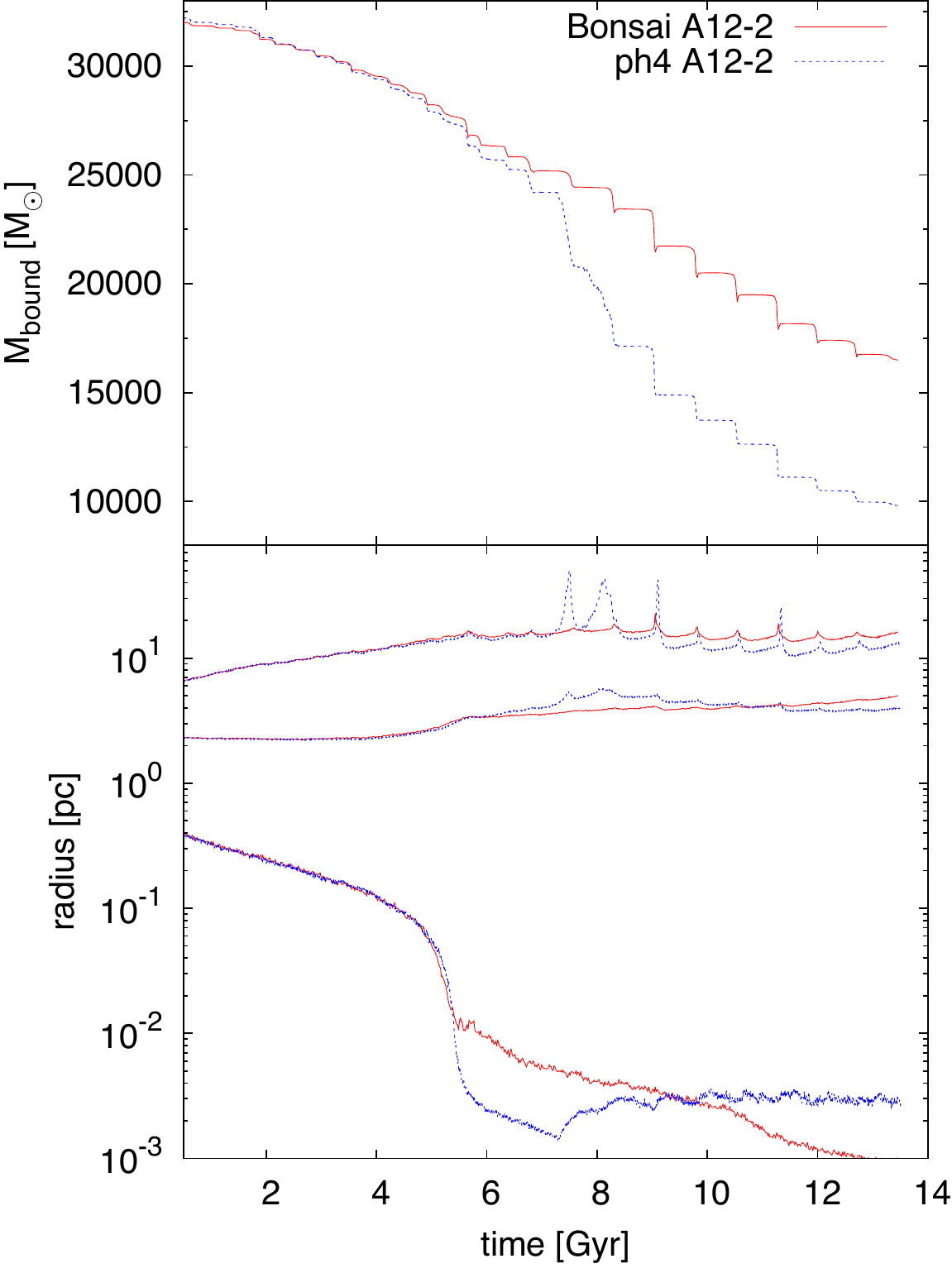}
  \caption[]{Evolution of the bound mass (top panel) and the 90\%,
    50\% and 1\% Lagrangian radii (top to bottom, bottom panel) of
    star cluster A12-2. The cluster was simulated with Bonsai (red,
  solid curves) and ph4 (green, dashed curve).}
  \label{Fig:TreeCodevsDirect}
\end{figure}

In order to validate the use of the tree code, we compare a cluster
simulated with Bonsai to a reference simulation using ph4.  In
Fig.\,\ref{Fig:TreeCodevsDirect}, we present the mass and the
Lagrangian radii of this simulated cluster as a function of time, for
both Bonsai and ph4. 

The difference in mass evolution between ph4 and Bonsai remains quite
small until about 5\,Gyr. After this moment, both clusters go into core
collapse and the two codes start to deviate more. Until about 8.5\,Gyr,
the ph4 cluster displays much higher mass loss than the Bonsai cluster
as it expands following core collapse. After 8.5\,Gyr, both codes again
show similar behaviour.

The Lagrangian radii of the clusters are nearly equal until core
collapse occurs at about 5\,Gyr. After this, the core collapse is
initially deeper in ph4, while after 8.5\,Gyr Bonsai reaches the same
depth.

From these results, we infer that our Bonsai simulations are not as
well suited for determining the internal structure and evolution of
the star clusters as ph4 would be, and underestimates mass loss due to
core collapse. However, the effect of the tidal field on the mass loss
rate is similar in ph4 and Bonsai. Since we investigate only the mass
evolution of the clusters due to the tidal field in this article, we
conclude that Bonsai gives an adequate indication of the effect of
tidal fields on the cluster mass loss, and that it can be used to
study the survivability of star clusters.

\subsection{The tidal field}

In each snapshot of the CosmoGrid simulation we calculate the tidal
tensor at the location of the selected dark-matter particle, which
represents a star cluster.  The contribution to the tidal tensor
$\mathbf{T}_{\rm t}$ from a particle with relative position
$\mathbf{r}$ is given by the second derivative of the gravitational
potential $\phi$: 
\begin{equation} T_t^{ij}(\mathbf{r'})=-\frac{\partial^2
  \phi}{\partial {r'^i \partial r'^j}} \end{equation} where
$\mathbf{r}' = \mathbf{r} + \epsilon$. For the CosmoGrid simulation,
the value for the softening length $\epsilon$ was 175\,parsec. 

The strength of the tidal field scales as  $\frac{\partial^2
\phi}{\partial^2 \mathbf{r}} \sim \frac{1}{r^3}.$ Any particle at a
distance of about $\epsilon$ will have as much effect on the tidal
tensor as the largest halo in the CosmoGrid simulation (containing
$\sim 3\times10^8$ particles) would have at a distance of $\sim
117$\,kpc. We therefore include the contribution from all particles
within a radius of 125\,kpc from our clusters to determine
$\mathbf{T}_{\rm t}$.

The strength of the tidal field is calculated from the eigenvalues
$\lambda_i$ and eigenvectors $\mathbf{\nu}_i$ of this tensor
$\mathbf{T}_{\rm t}$. The eigenvalues give the magnitude of tidal
field, whereas the eigenvectors give the direction along which the
system is stretched.  This method for calculating the tidal tensor is
similar to the one employed in \cite{2011MNRAS.418..759R}. 

Since we calculate the tidal tensor $\mathbf{T}_{\rm t}$ from snapshots of
the CosmoGrid simulation and the number of snapshots is limited, we do
not have a continuous tidal field. In order to prevent large, sudden
changes, we linearly interpolate the tidal tensor between snapshots to
create a continuous tidal tensor. In Sec.\,\ref{Sec:Validation}, we
validate this interpolation method.

\subsection{Combining the clusters and the tidal field}

We use the tidal tensor to calculate the external potential acting on
each of the stars in the simulated clusters. We integrate the internal
potential of the clusters with this external potential using a
Bridge-like scheme \citep{2007PASJ...59.1095F}, which is implemented
in AMUSE \citep{2012arXiv1204.5522P}. 

This scheme can be used to combine interacting systems that
are calculated in different instances and/or using different codes,
i.e. multiple interacting star clusters, globular clusters in a
galactic environment, a galactic disk in a halo potential or embedded
star clusters \citep{2011MNRAS.tmp.2133P}.

In this scheme, the cluster experiences the gravity from the external
field through periodic velocity kicks. It alternates between these
velocity kicks and a drift due to self gravity evolution of the
system. During one time step, the system first experiences a kick of
the velocities over a time step $dt/2$, then a drift over a time step
$dt$, and finally another kick over $dt/2$. In our setup, the external
potential is derived from the CosmoGrid simulation and therefore
necessarily fixed, while the clusters receive velocity kicks from the
external tidal field.

\subsection{Escaping and bound stars}

In the simulations, we calculate a tidal radius from the cluster mass
and the strength of the tidal field (given by the largest eigenvalue
of the tidal tensor $\lambda_{\rm max}$). This tidal radius is equal to 
\begin{equation}
  R_{\rm tidal} = \left(\frac{G M}{\lambda_{\rm max}}\right)^{1/3}.
\end{equation}
Particles at a distance from the cluster centre larger than $R_{\rm
tidal}$ will experience a larger force from the external tidal field
than from the cluster's own internal mass. It is then considered an
`escaping particle', and not included in the cluster's bound mass
(defined as the total mass inside $R_{\rm tidal}$). If the particle
returns to a position within the tidal radius, this is reversed. If it
moves to a distance $> 10 R_{\rm tidal}$, the particle is removed from
the simulation.

At times when the cluster is located in the centre of a local
subconcentration of dark matter or near a halocentre, the value of
$\lambda_{\rm max}$ may become negative for a short period. In such
cases, the tidal radius is not defined. At such times, the bound mass
of the cluster is not evaluated and no mass loss is experienced by the
cluster. Stars that may have escaped from the cluster during this
period will however still be removed once the tidal radius is again
defined.

\subsection{Validation}\label{Sec:Validation}

We validate the simulation environment by comparing our results with
those obtained using a different method and to a star cluster in
isolation.  As a reference model, we simulate a star cluster in an
orbit with $R_{\rm apo} = 12\hkpc$ and $e = 0.71$ around a point-mass
of $10^{10}$\,\Msun, by including this point-mass in the simulation.
The simulated cluster contains 8\,000 equal-mass stars of 1\,\Msun
within a \cite{1911MNRAS..71..460P} sphere with a virial radius of
3\,pc.  We compare this model to a simulation where we first
calculated the tidal field that would be experienced by such a cluster
and used this as an external field for the simulation using the Bridge
scheme.  For this test, we run simulations with ph4, as the large
difference in particle masses would make the reference simulation
unsuitable for a single-precision tree code like Bonsai.

\begin{figure}
  \includegraphics[width=\columnwidth]{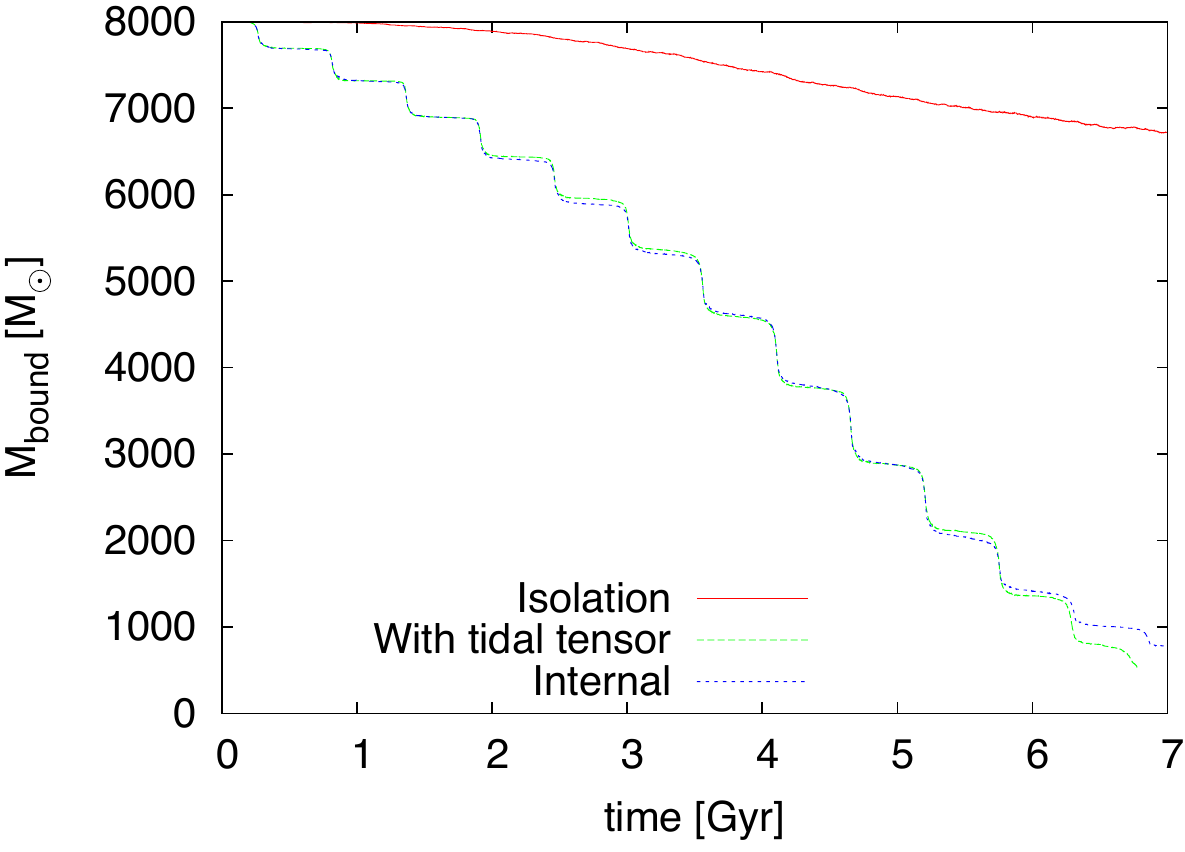}
  \caption[]{Mass evolution of a 8\,000 star cluster in isolation
  (solid curve), and in orbit around a point mass of $10^{10}$\,\Msun
  (dashed and dotted curves). The dashed curve gives the
  mass-evolution of the cluster when the tidal field is incorporated
  in the gravitational $N$-body simulation using Bridge. The tidal
  field in the latter case was resolved at the resolution of the
  $N$-body integrator. The dotted curve gives the mass-evolution when
  the point mass is directly included in the simulation.}
\label{Fig:ValidationMethod}
\end{figure}

The mass evolution of these test simulations are presented in
Fig.\,\ref{Fig:ValidationMethod}. The difference in the mass evolution
of the cluster in the simulations with and without Bridge is quite
small, indicating the validity of this method for this set of
parameters.

Also, we validate the effect of the discretisation at which the tidal
tensor is evaluated. Since the time-resolution of the cosmological
simulation is limited to about 35\,Myr at high $z$ and 17.5\,Myr at
low $z$, anomalies will be visible for clusters with an orbital period
of this order. Ideally, one would like to increase the number of
snapshots for the cosmological simulation in order to obtain a higher
time-resolution for the tidal tensor. 

\begin{figure}
  \includegraphics[width=\columnwidth]{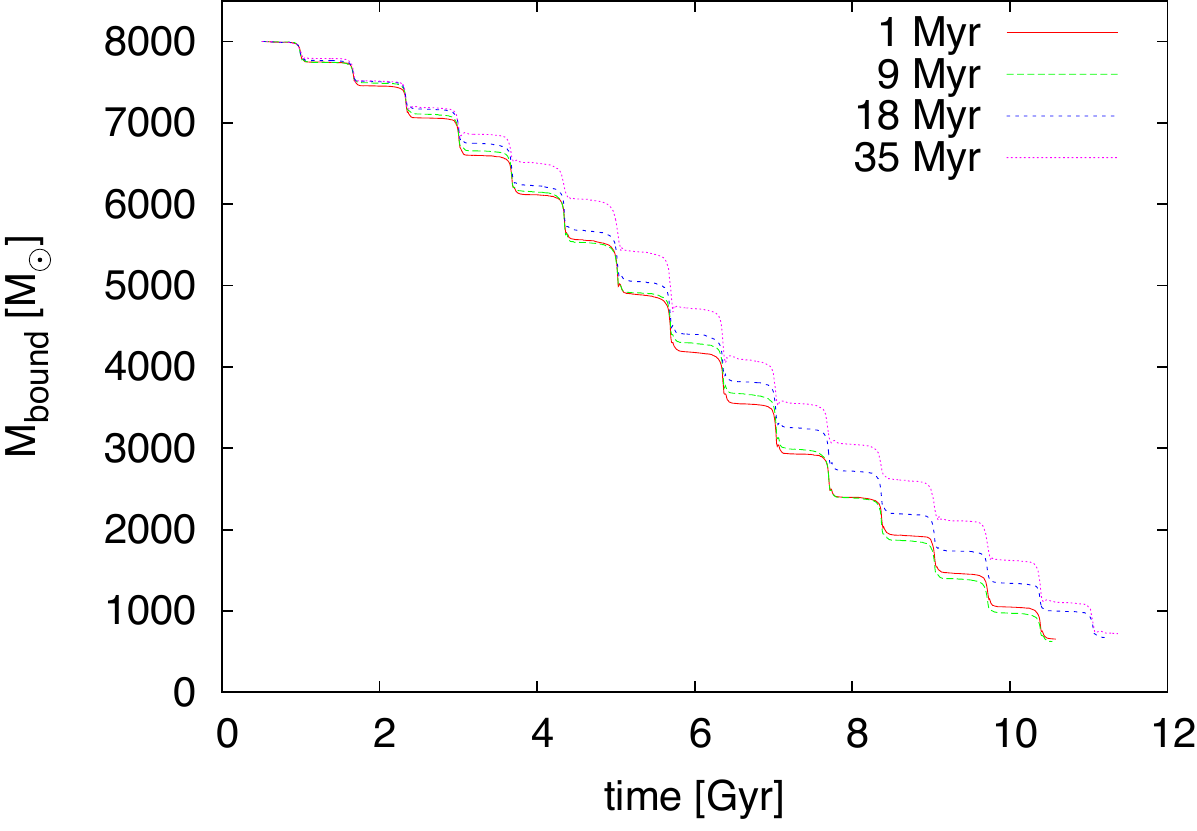}
  \caption[]{Mass evolution of a 8\,000 star cluster with tidal field
    calculated from a halo potential using the tidal tensor and
    evaluated using Bridge at discrete time intervals. The cluster
    orbits in the halo potential with $R_{\rm apo} = 15\,{\rm kpc}$
    and $e = 0.85$. In order to study the effect of the time
    resolution of the tidal field, we evaluate the tidal tensor with
    time intervals of 1, 9, 18 and 35\,Myr.}
\label{Fig:ValidationTimestepping}
\end{figure}

In Fig.\,\ref{Fig:ValidationTimestepping} we present the mass
evolution of a cluster in a static halo potential
\citep{1990ApJ...348..485P} with a core mass of $10^{9}$\,\Msun\,and a
core radius of $1$\,kpc, now using tidal tensors sampled with a time
resolution ranging from 1 to 35\,Myr. The orbital parameters of the
cluster in this potential are $R_{\rm apo}=15$\,kpc, $e=0.85$, and the
cluster orbits the potential in $670$\,Myr. The mass loss rates for the
clusters using a tidal tensor sampled with a time interval of 1\,Myr
and 9\,Myr have converged in this orbit, while the clusters using tensors
sampled with a 17.5 or 35\,Myr time interval show a slightly
reduced mass loss rate and therefore longer lifetimes. For clusters
with closer orbits, this effect will be stronger. In a cosmological
setting, any change in the potential that takes place on a timescale
similar to or smaller than the sampling rate cannot be taken into
account properly, and may also lead to errors in the mass loss rate.

\section{Results}

\subsection{The evolution of the selected haloes}

We perform simulations of star clusters in two selected haloes, which
we call halo A and halo B. Halo A experiences two major merger events,
the last of which is not completed by $z=0$ (see
Figs.\ref{Fig:MergerHistory} and \ref{Fig:HaloMassEvolution}), but
otherwise its evolution is rather gradual from $z=65$ to $z=0$. By the
end of the simulation (at $z=0$) its mass is $6.3\times
10^{11}\,\hMsun$.

Halo B has a more violent history; it experiences two major mergers,
one between 4\,Gyr and 5\,Gyr and one with comparable mass between
8\,Gyr and 10\,Gyr (see also Figs.\,\ref{Fig:MergerHistory} and
\ref{Fig:HaloMassEvolution}).  By the end of the simulation at $z=0$
it has a mass of $4.8 \times 10^{11}$\,\hMsun\, (see
Tab.\,\ref{Tab:HaloProperties}). Even though the total mass of halo B
is slightly smaller than that of halo A, both haloes contain a mass of
$7\times10^{9}$\,\hMsun\, within 3\,kpc from the halo centre (see
Fig.\,\ref{Fig:RadialProfiles}).

In each halo we have selected 15 dark-matter particles which are
initialized at $z=10$ as star clusters, and evolved with the
cosmological simulation as a background potential.

\subsection{The evolution of the star clusters}

We perform 30 simulations of star clusters with a tidal field; 15 are
initialized in halo A and 15 in halo B. An additional cluster is
simulated in isolation, to identify the mass loss component caused by
relaxation. For this cluster, a radius of 200\,pc is used to determine
if stars are bound. All clusters are born at $z=10$ (corresponding to
an age of the Universe of about 500\,Myr) with a total mass of $\sim
32\,000$\,\Msun\, and an initial virial radius of 3 parsec for each
cluster.  All stars have the same mass and we did not include stellar
evolution.

\begin{figure}
  \includegraphics[width=\columnwidth]{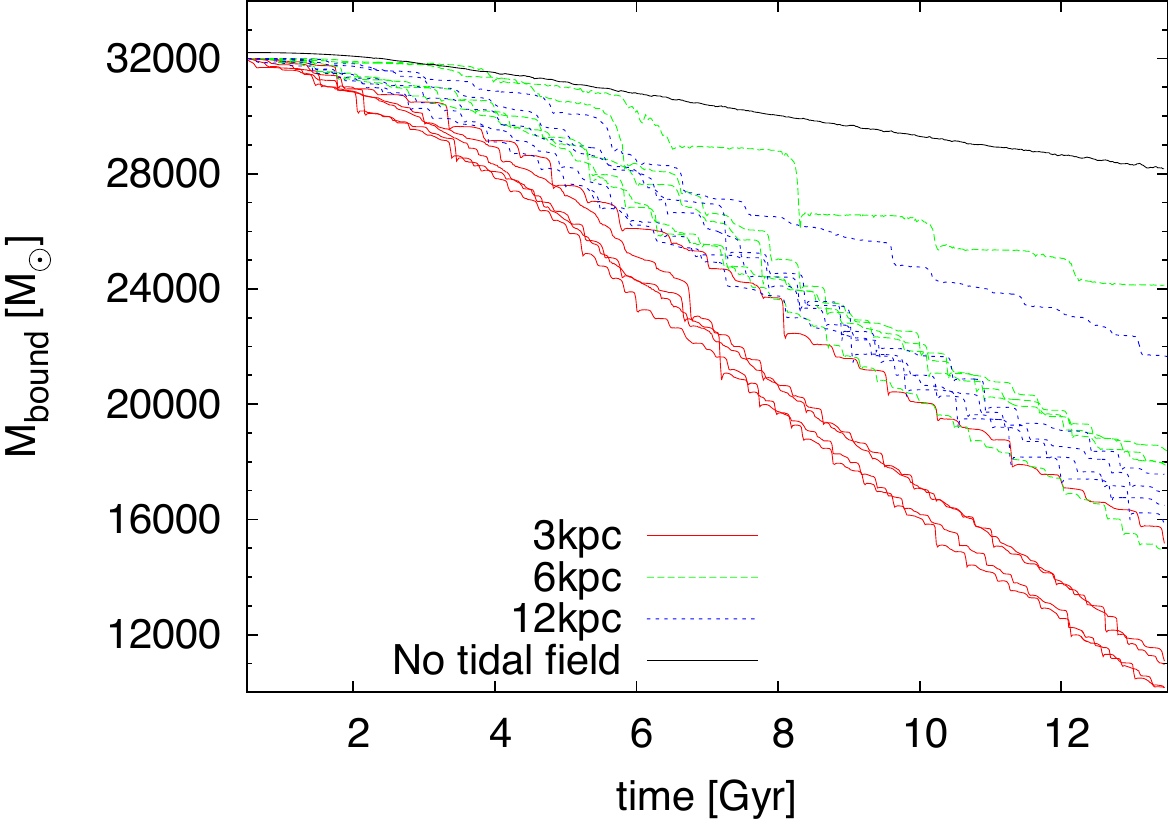}\\
  \includegraphics[width=\columnwidth]{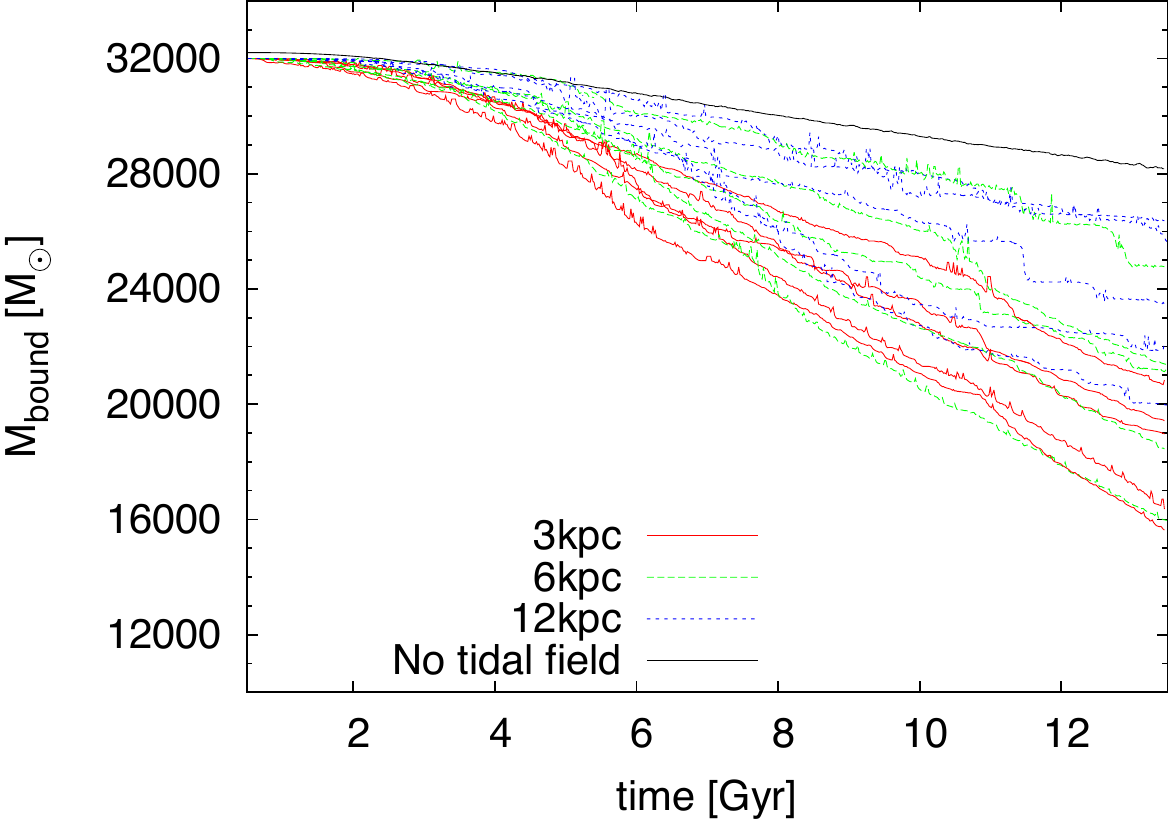}
  \caption[]{Bound mass in the simulated clusters for haloes A (top
  figure) and B (bottom figure). The red, solid; green, dashed and
  blue, dotted lines indicate clusters selected at $3$, $6$ and
  $12\hkpc$, respectively. The black line indicates the mass of a
  cluster without an external tidal field.
  
  } \label{Fig:BonsaiClustermass}
\end{figure}

In Fig.\,\ref{Fig:BonsaiClustermass} and Tab.\,\ref{Tab:BonsaiRuns} we
present the mass evolution and final masses of these simulations. The
mass evolution of all simulated clusters is rather gradual
irrespective of the sudden events in the growth of the host haloes.
The small and rather sudden changes in mass are caused by the
pericentre passages of the clusters in its orbit around the dark
matter host. The clusters with a smaller orbital separation at z=0
tend to lose mass at a higher rate.

The averaged final mass of the star clusters in halo A is smaller
than the final mass of those that evolved in halo B for each of the
radial bins (see Tab.\,\ref{Tab:BonsaiRuns}).  This is consistent with
halo B originating from a larger number of less massive haloes,
causing the tidal forces experienced by clusters in this halo over
time to be smaller.  For both haloes the clusters selected around
$3\hkpc$ show the strongest mass loss; this is noticeable from the
first few Gyr on. This behaviour is as expected from the distribution
of particles at $z=10$ (see Fig.\,\ref{Fig:HaloZ10Globulars}), where
we see that the particles that end up in the central parts are already
more concentrated at high redshift.  The rate of mass loss for these
clusters proceeds more gradual compared to the clusters in wider
orbits.  Integrated over time clusters lose mass at a rather constant
rate.

\begin{table}
  \centering
  \caption{Results for star cluster simulations using Bonsai, with
  various tidal tensors. Each cluster contains 32\,000 equal mass
  stars of about 1\Msun, and has an initial radius of 3 parsec. We
  use a softening length of $40/N$. Cluster type is Immigrant (I) or
  Native (N). The clusters indicated with a bold font are displayed in
  Figs.\ref{Fig:HaloAClusters} and \ref{Fig:HaloBClusters}.}
  \label{Tab:BonsaiRuns} \begin{tabular}{l|l|l|l|l}
    Cluster ID & Halo & Type & Distance at z=0 & Mass at z=0\\
    \hline
    \bf{Isolated}&    &      &                   & \bf{28174 \bfMsun} \\
    \hline
    \hline
    A3-1 & A    & I    & $3 \hkpc$         & 10157 \Msun \\
    A3-2 & A    & I    & $3 \hkpc$         & 15171 \Msun \\
    A3-3 & A    & N    & $3 \hkpc$         & 10969 \Msun \\
    A3-4 & A    & N    & $3 \hkpc$         & 11085 \Msun \\
    A3-5 & A    & N    & $3 \hkpc$         & 10169 \Msun \\
    \bf{Average}&     &      &                   & \bf{11510 \bfMsun} \\
    \hline
{\bf A6-1}&A    & I    & $6 \hkpc$         & 17906 \Msun \\
    A6-2 & A    & N    & $6 \hkpc$         & 24130 \Msun \\
    A6-3 & A    & N    & $6 \hkpc$         & 14964 \Msun \\
{\bf A6-4}&A    & N    & $6 \hkpc$         & 17978 \Msun \\
    A6-5 & A    & N    & $6 \hkpc$         & 18430 \Msun \\
    \bf{Average}&     &      &                   & \bf{18682 \bfMsun} \\
    \hline
    A12-1 & A    & I    & $12 \hkpc$        & 21677 \Msun \\
    A12-2 & A    & I    & $12 \hkpc$        & 16487 \Msun \\
    A12-3 & A    & N    & $12 \hkpc$        & 17574 \Msun \\
    A12-4 & A    & N    & $12 \hkpc$        & 15873 \Msun \\
    A12-5 & A    & N    & $12 \hkpc$        & 16904 \Msun \\
    \bf{Average}&     &      &                   & \bf{17703 \bfMsun} \\
    \hline 
    \hline
    B3-1 & B    & I    & $3 \hkpc$         & 19422 \Msun \\
    B3-2 & B    & N    & $3 \hkpc$         & 18974 \Msun \\
    B3-3 & B    & N    & $3 \hkpc$         & 16366 \Msun \\
    B3-4 & B    & N    & $3 \hkpc$         & 15633 \Msun \\
    B3-5 & B    & N    & $3 \hkpc$         & 20839 \Msun \\
    \bf{Average}&     &      &                   & \bf{18247 \bfMsun} \\
    \hline
    B6-1 & B    & I    & $6 \hkpc$         & 24774 \Msun \\
{\bf B6-2}&B    & N    & $6 \hkpc$         & 18446 \Msun \\
{\bf B6-3}&B    & I    & $6 \hkpc$         & 15988 \Msun \\
    B6-4 & B    & N    & $6 \hkpc$         & 21128 \Msun \\
    B6-5 & B    & N    & $6 \hkpc$         & 21398 \Msun \\
    \bf{Average}&     &      &                   & \bf{20347 \bfMsun} \\
    \hline
    B12-1 & B    & I    & $12 \hkpc$        & 19986 \Msun \\
    B12-2 & B    & I    & $12 \hkpc$        & 23479 \Msun \\
    B12-3 & B    & N    & $12 \hkpc$        & 25819 \Msun \\
    B12-4 & B    & N    & $12 \hkpc$        & 26374 \Msun \\
    B12-5 & B    & I    & $12 \hkpc$        & 22137 \Msun \\
    \bf{Average}&     &      &                   & \bf{23559 \bfMsun} \\
    \hline

  \end{tabular}
\end{table}

\begin{figure*}
  \subfloat[]{
    \includegraphics[width=\columnwidth]{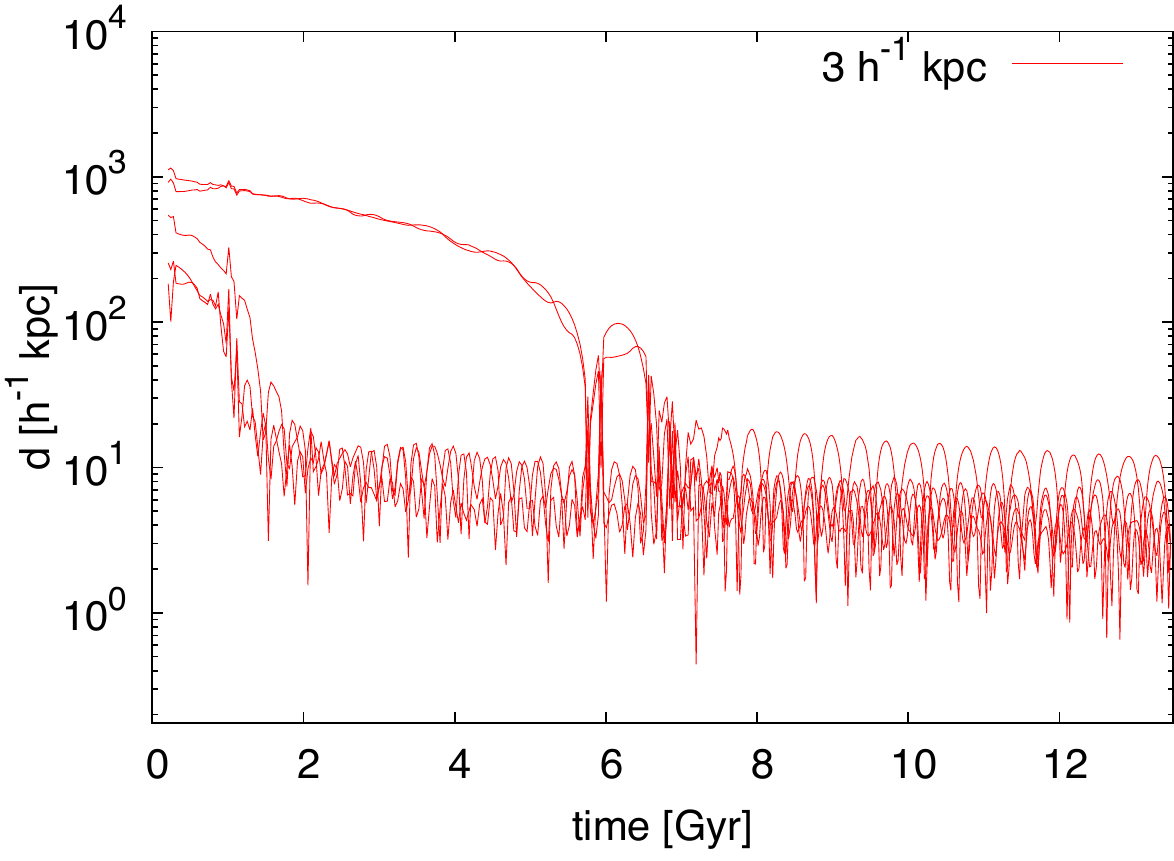}
  }
  \subfloat[]{
    \includegraphics[width=\columnwidth]{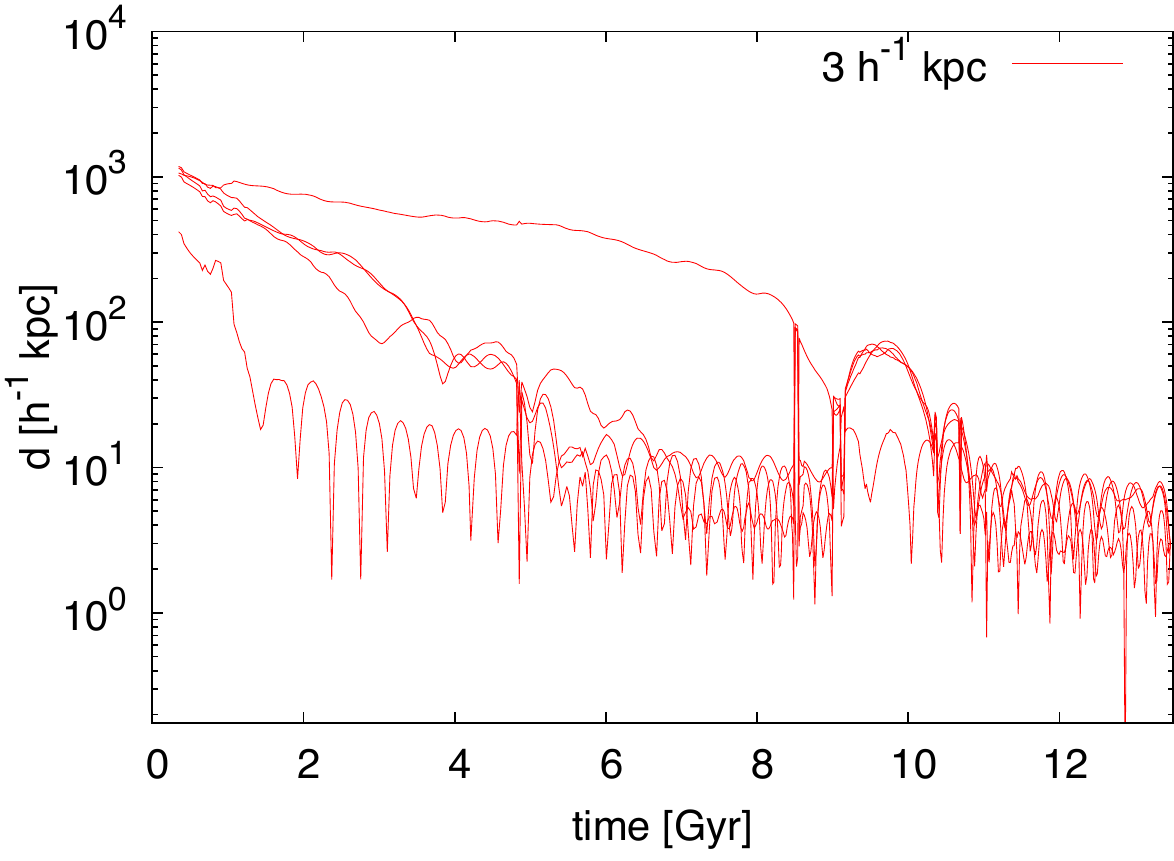}
  }

  \subfloat[]{
    \includegraphics[width=\columnwidth]{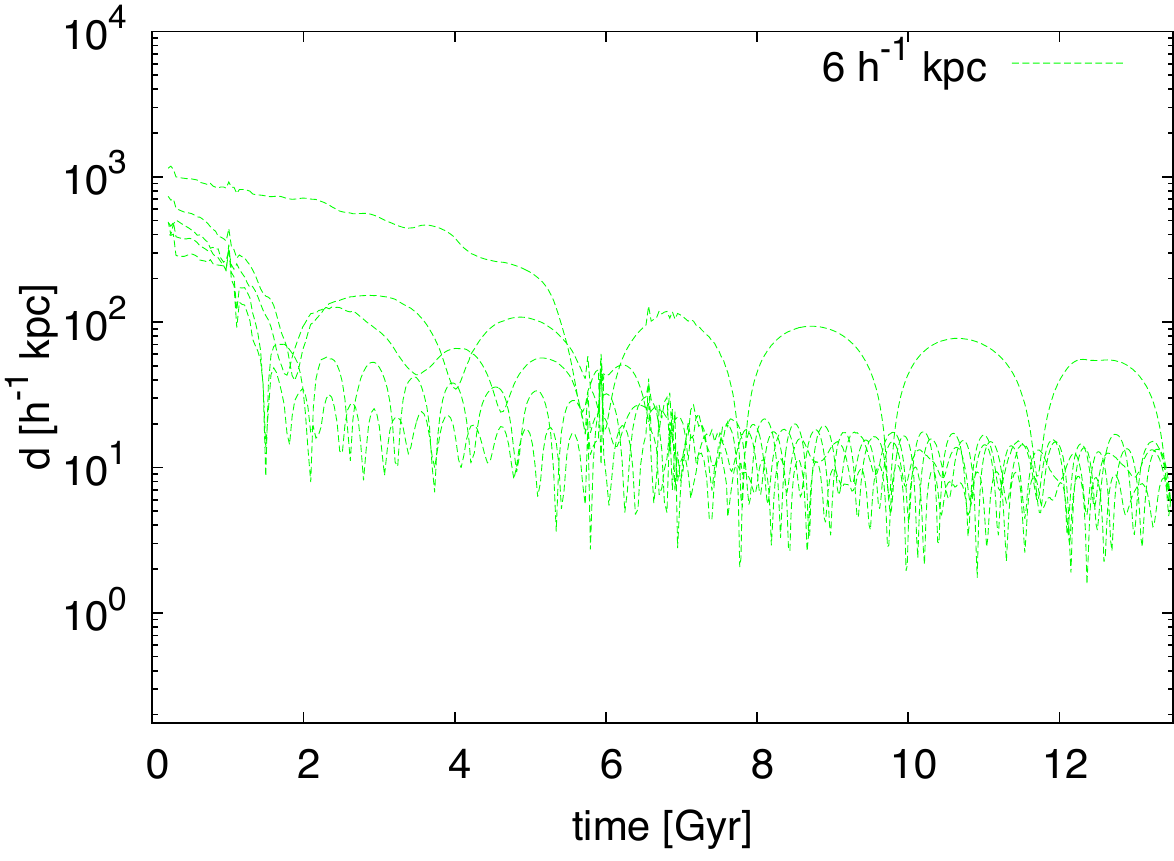}
  }
  \subfloat[]{
    \includegraphics[width=\columnwidth]{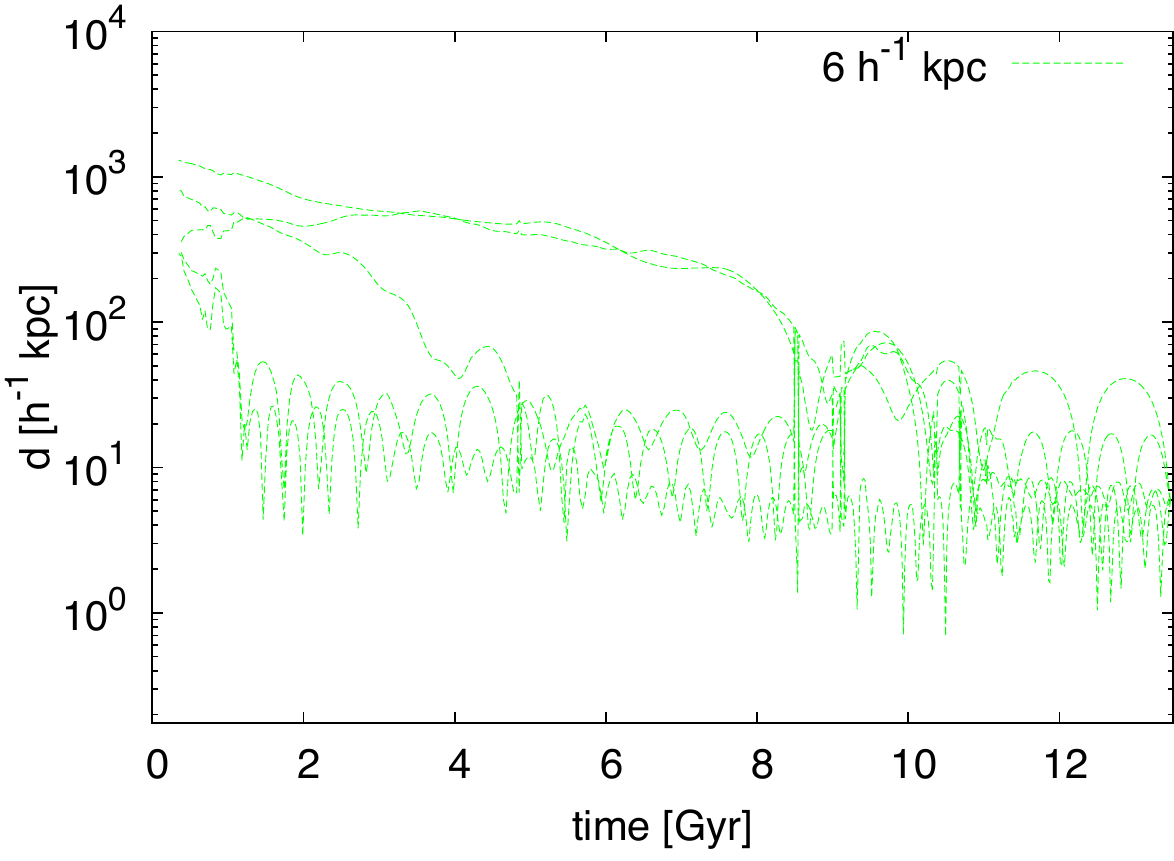}
  }

  \subfloat[]{
    \includegraphics[width=\columnwidth]{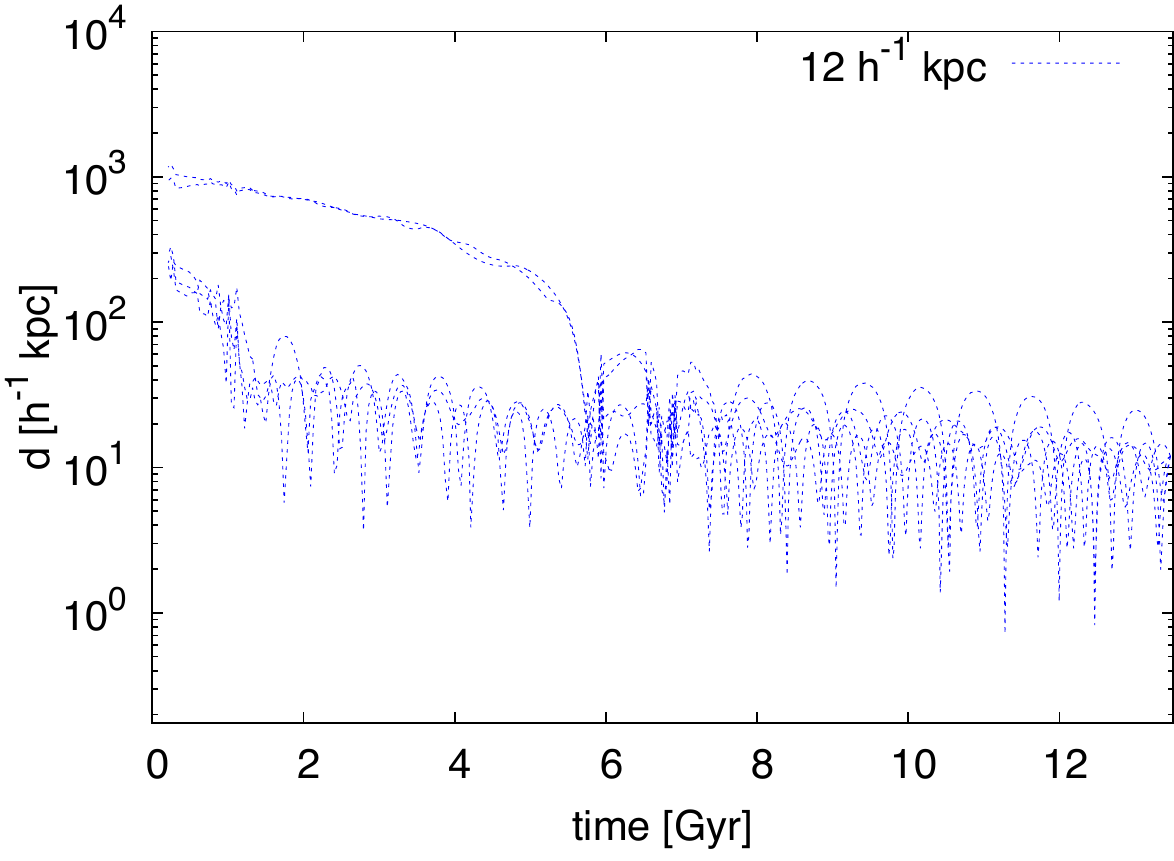}
  }
  \subfloat[]{
    \includegraphics[width=\columnwidth]{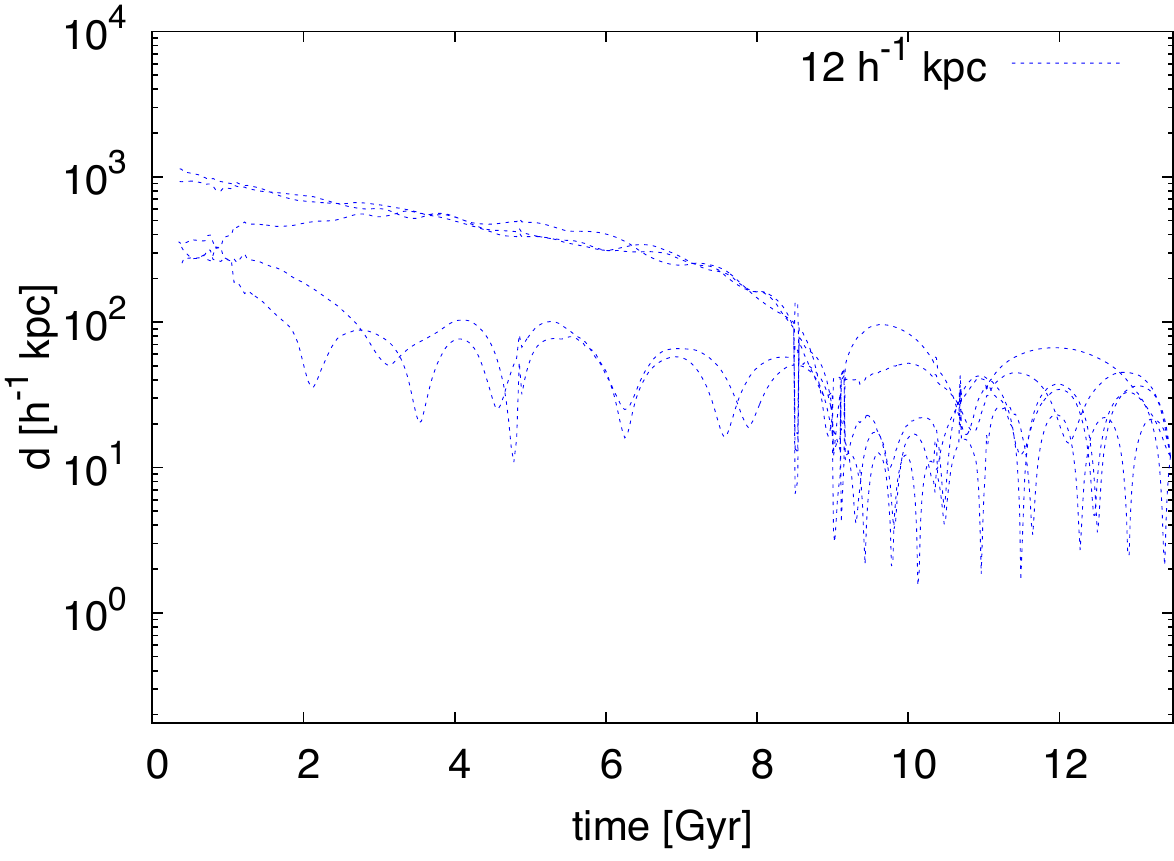}
  }
  \caption[]{Co-moving distance of the clusters to the main halo
    centre for clusters in halo A (left) and halo B (right). Top,
    middle and bottom figures show clusters selected at $3$, $6$ and
    $12\hkpc$, respectively.  Merger events in both haloes are visible
    as particles fall towards the halo centre. The orbital periods of
    the clusters are clearly visible.}
  \label{Fig:TimevsDistance}
\end{figure*}

In Tab.\,\ref{Tab:BonsaiRuns} we distinguish between two types of
clusters: those that were part of the main halo before the final
completed major merger event (at $t=6$\,Gyr and $t=8$\,Gyr for haloes
A and B, respectively), and those that are accreted upon or after this
event.  The former we identify with `native' clusters, and the others
as `immigrants'.  The difference between immigrant clusters and native
clusters is apparent in the Figs.\,\ref{Fig:TimevsDistance}.

In Fig.\,\ref{Fig:TimevsDistance} we present the orbital evolution of
the selected dark-matter particles (i.e.\, the clusters) from halo A
(left) and halo B (right).  In the following paragraphs, we discuss the
evolution of two clusters from each halo; one immigrant and one native
cluster in more detail.

\subsubsection{The clusters in halo A}

\begin{figure*}
  \subfloat[]{
    \includegraphics[width=\columnwidth]{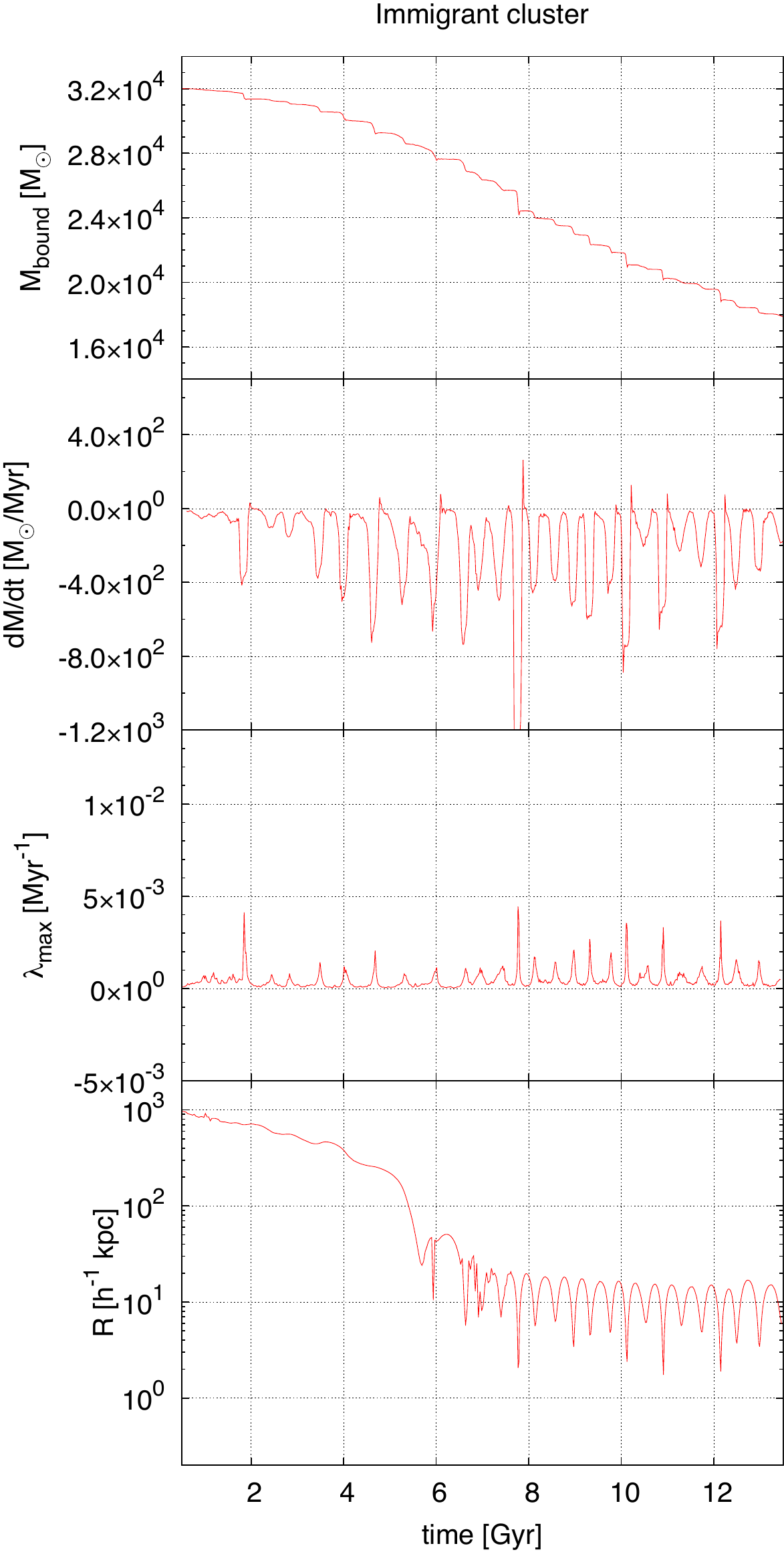}
  }
  \subfloat[]{
    \includegraphics[width=\columnwidth]{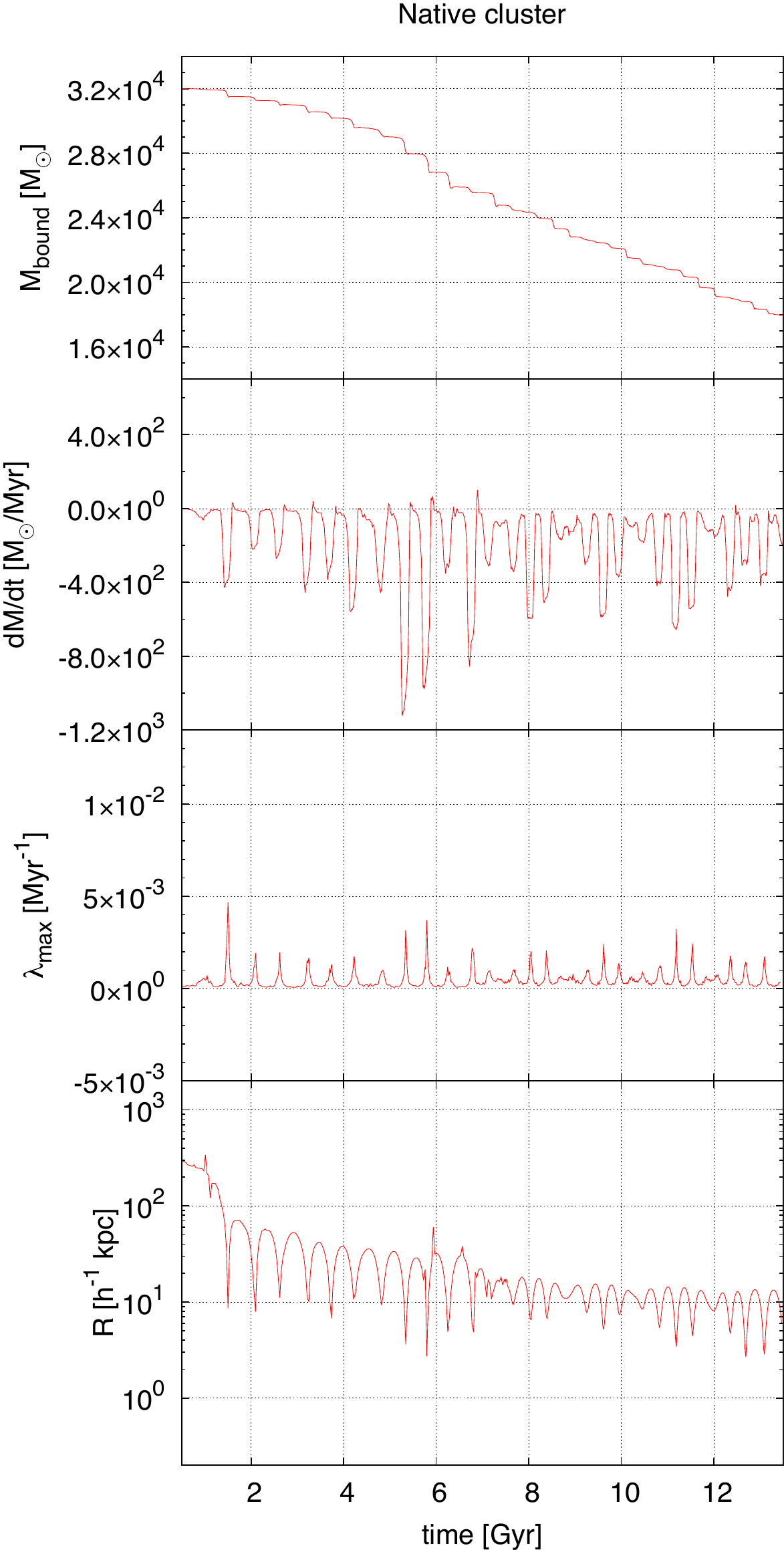}
  }
  \caption[]{Evolution of two star clusters in orbit around halo A.
    To the left is a typical immigrant cluster identified with
    dark-matter particle nr A6-1 and to the right we present a
    native cluster, particle nr A6-4 (see also
    Tab.\ref{Tab:BonsaiRuns}).  From top to bottom the panels show the
    bound mass, the mass-loss rate, the strength of the tidal field
    ($\lambda_{\rm max}$) and the co-moving distance of the cluster to
    the centre of the dark matter halo. The decrease seen in the
    co-moving distance, apart from the merger event, is caused by the
    expanding cosmic volume. \label{Fig:HaloAClusters}}
\end{figure*}

In Fig.\,\ref{Fig:HaloAClusters}, we show the mass evolution, mass
loss rate over intervals of 10 Myrs and tidal field strength for an
immigrant cluster (nr.\, A6-1, left) and a native cluster (nr.\,
A6-4, right).  This halo experiences a major merger which starts
at about 6\,Gyr, at that time the immigrant cluster is also captured
by the main halo. The merger lasts until about 8\,Gyr (see also
Fig.\,\ref{Fig:MergerHistory}). 
When the merger is finished, the tidal field strength $\lambda_{\rm
max}$ shows more frequent peaks, indicating it has a shorter orbital
period than before the merger. However, there is little change in the
mass-loss rate.

The native cluster of halo A experiences the same merger but was
already member of the major halo. Its orbit becomes somewhat less
eccentric after the merger, while its apocentric distance and orbital
period decrease. The mass-loss rate from this cluster is mostly
unaffected by the merger.

\begin{figure}
  \includegraphics[width=\columnwidth]{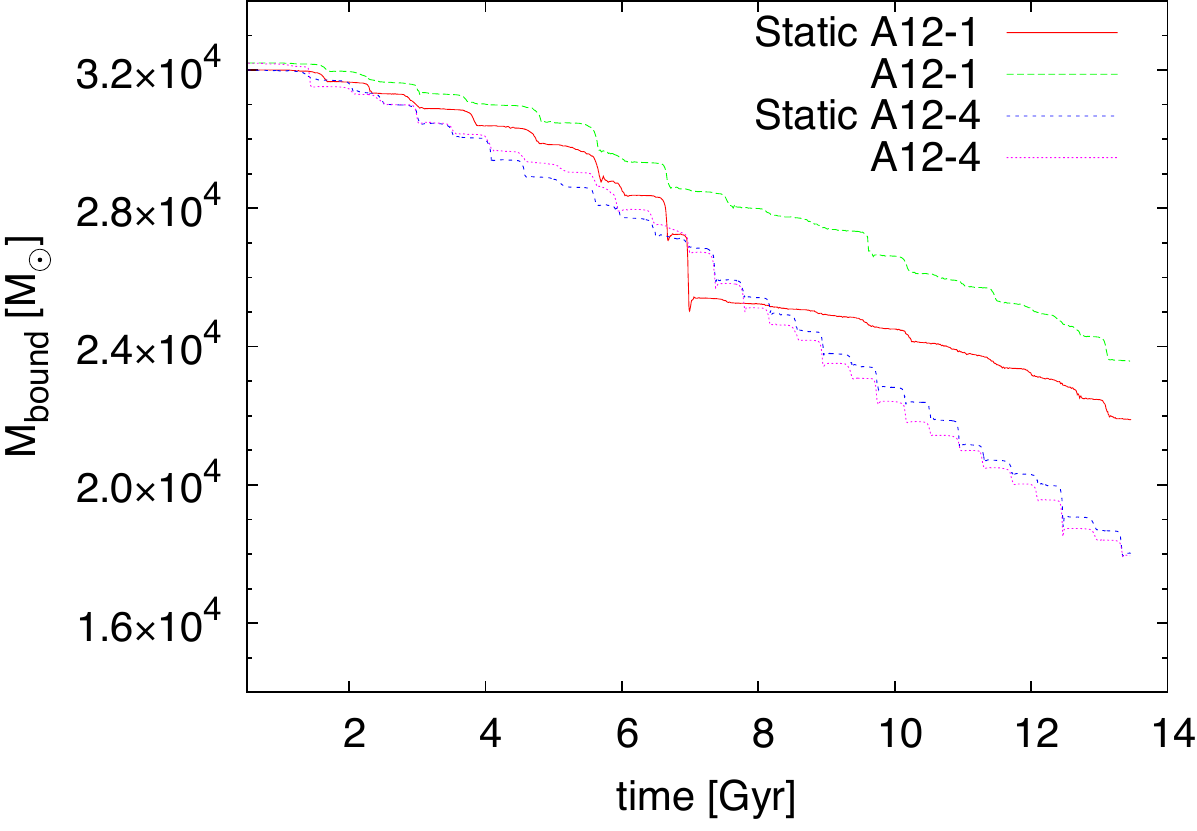}
  \caption[]{Evolution of two star clusters, A12-1 (immigrant) and
    A12-4 (native). Both are simulated in an evolving halo as well as
    a static halo.  \label{Fig:StaticvsEvolving}}
\end{figure}

In Fig.\,\ref{Fig:StaticvsEvolving}, we show the result of two star
clusters (A12-1 and A12-4), both using the tidal tensor calculated
from the evolving CosmoGrid halo and using a tidal tensor calculated
from the static $z=0$ CosmoGrid halo. For the static halo case, we
sampled the tensor using the orbital trajectory of the cluster around
its parent halo. For the native cluster, the resulting mass evolution
differs only marginally, while the immigrant cluster suffers
considerably higher mass loss in the static halo case, especially
around the time the halo merger takes place, when its orbit is
erratic. The static halo is a good approximation for the native
cluster, while it falls short for the immigrant cluster.

\subsubsection{The clusters in halo B}

\begin{figure*}
  \subfloat[]{
    \includegraphics[width=\columnwidth]{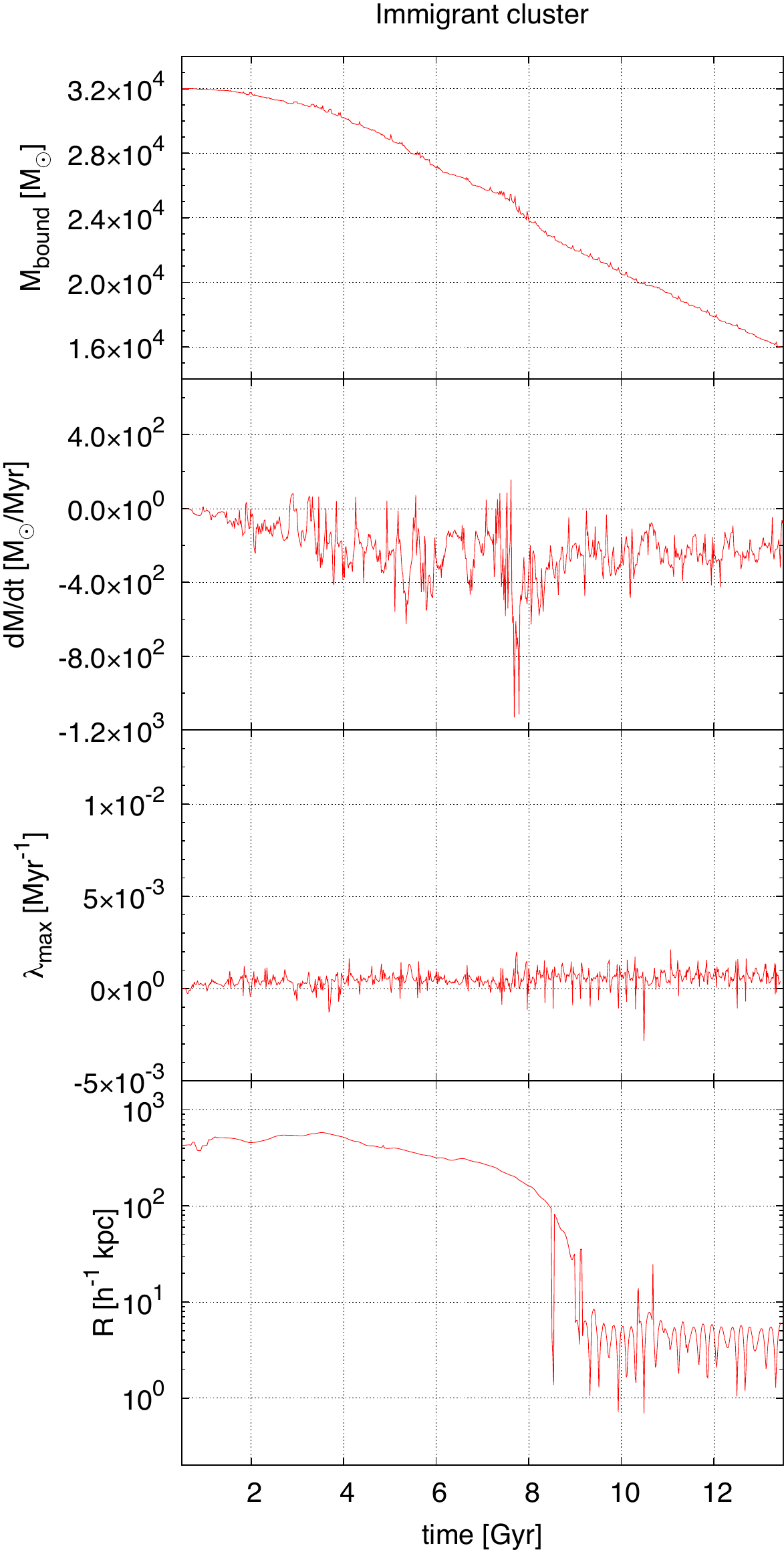}
  }
  \subfloat[]{
    \includegraphics[width=\columnwidth]{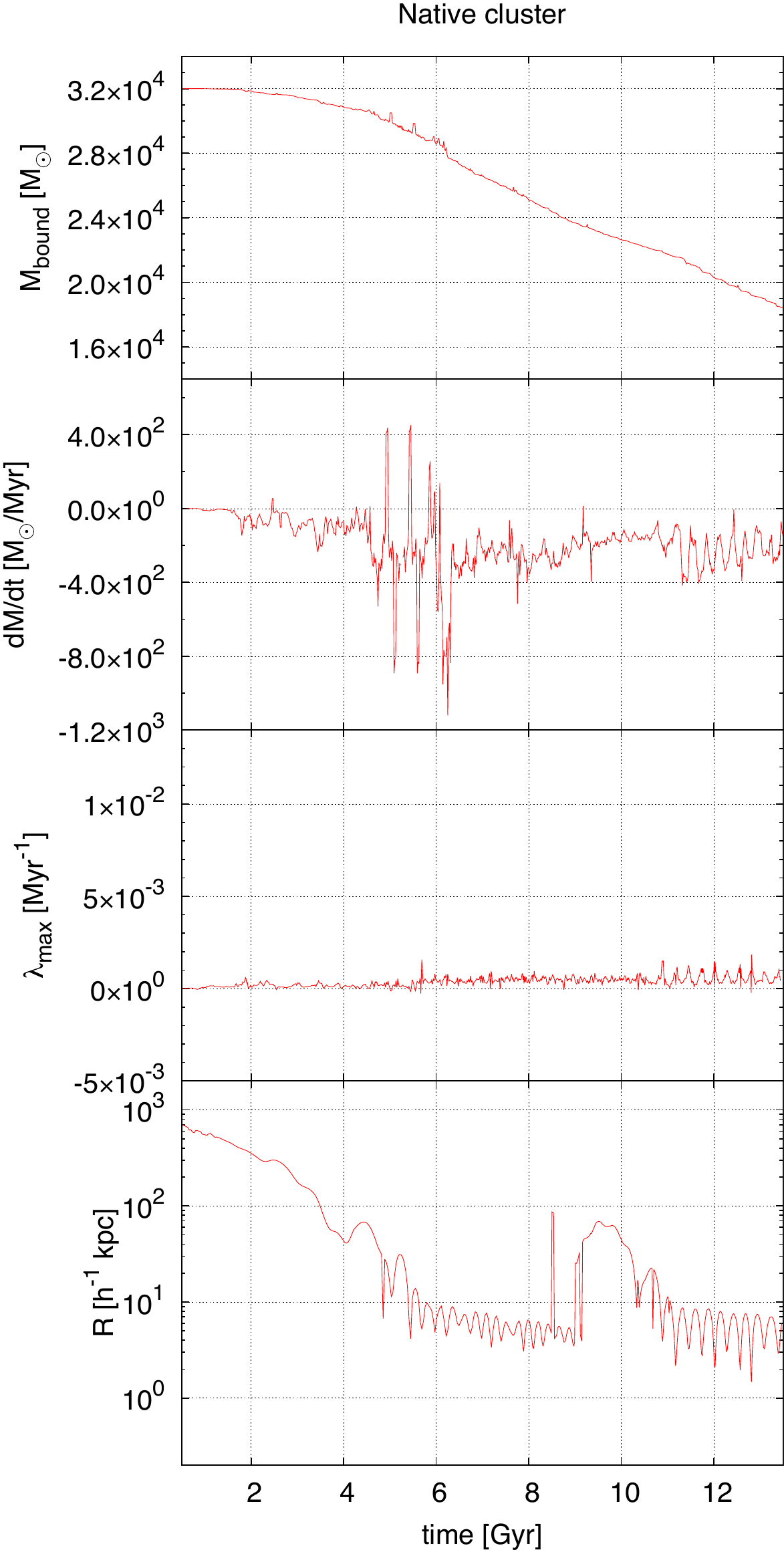}
  }
  \caption[]{Evolution of two star clusters in orbit around halo B.
    To the left is a typical immigrant cluster identified with
    dark-matter particle nr B6-3 and to the right we present a
    native cluster, particle nr B6-2 (see also
    Tab.\ref{Tab:BonsaiRuns}).  From top to bottom the panels show the
    bound mass, the mass-loss rate, the strength of the tidal field
    ($\lambda_{\rm max}$) and the co-moving distance of the cluster to the
    centre of the dark matter halo. Negative values of $\lambda_{\rm
    max}$ correspond to times where the cluster is located inside a
    local (sub)concentration of matter.  The spikes in the bottom
    panels for the immigrant and the native clusters is associated
    with a confusion in identifying the main parent in the halo
    finder.
  \label{Fig:HaloBClusters}}
\end{figure*}

In Fig.\,\ref{Fig:HaloBClusters} we present the details of the
evolution of an immigrant cluster (nr.\,B6-3, left) and a native
cluster (nr.\,B6-2, right) of halo B.  The immigrant cluster is
captured during the major merger event that starts around $t=8$\,Gyr.
Different from the clusters in halo A, the orbital period of the
cluster around the dark matter halo is hardly visible in the cluster's
mass evolution (see Fig.\,\ref{Fig:HaloBClusters}, left, top panel). Just
before the merger (around $t=7$\,Gyr) the mass-loss rate is slightly
smaller than before or after the merger. This is caused by the
distortion of the infalling halo of which this cluster is a member at
that time.  When the merger is completed the cluster mass-loss rate
has resumed to be as high as before the merger (see
Fig.\,\ref{Fig:HaloBClusters}, left, second panel).

The native cluster becomes part of the main halo during its first
major merger, at around $t=5$\,Gyr. After this merger, the tidal forces
experienced by the cluster are stronger than before, visible in
Fig.\,\ref{Fig:HaloBClusters} (right, third panel) as a sudden
increase of $\lambda_{\rm max}$ after $t=6$\,Gyr. Its mass-loss rate
is also increased, as can be seen from Fig.\,\ref{Fig:HaloBClusters}
(right, second panel) and the difference in slope of $M_{\rm
bound}(t)$ at $t=4$\,Gyr and $t=8$\,Gyr.
The second major merger event leads to a temporary reduction in
mass-loss for the cluster (at around $t=10$\,Gyr). After this second
halo merger the orbital period is evidently visible in the mass
evolution of the cluster.

\section{Discussion and conclusions}

We presented a method to simulate star clusters within a
pre-calculated tidal field, using the AMUSE environment. As a proof of
concept, we apply this method to calculate the mass loss rates for
star clusters in two live \LCDM haloes from the CosmoGrid simulation.
Our method compares well to self-consistent simulations.

We find that the mass loss rate strongly depends on the cluster's
orbital parameters around the halo centre, as well as the central mass
of the halo.  Also, tidal disruption due to the \LCDM environment is
weaker in haloes that experienced many mergers. Finally, we find that
in a Milky Way-like halo, the contribution of the \LCDM environment to
$\Delta M/M_{\rm init}$ can be up to 0.6 for clusters with an initial
mass of 32\,000\Msun.

In this article, several effects have not been taken into account,
most notably we used a dark matter-only simulation. In this section,
we discuss the relevance of these effects and how we intend to address
these shortcomings.

Since the CosmoGrid simulation is a dark matter-only simulation, we do
not account for the effect of baryons. Because the formation of the
large-scale environment is dominated by dark matter, the lack of
baryons has little influence on the formation of the haloes. Star
clusters however likely contain little or no dark matter
\citep{2010IAUS..266..365B, 2011ApJ...741...72C}.

Because our simulation lacks baryons, there is no indication where
star clusters would have formed or ended up had they been formed in our
cosmological environment. We therefore have to resort to our method of
identifying tracer particles for the star clusters in the final
snapshot. We selected particles based on their distance to the halo
centre. However, it is very likely that globular clusters would have
formed in more specific locations, and possibly followed paths quite
different from the ones in our simulation. In a simulation that
includes baryonic matter and star formation, it would be possible to
detect locations and masses of star clusters as they form. 

The difference between old and young clusters herein is large: old
clusters (like the globular clusters in the Milky Way) formed before
there was a Galactic disk, and remain relatively free of its
influence. Young clusters however form in the galactic disk, and the
tides experienced by these clusters are dominated by encounters with
giant molecular clouds and spiral arms \citep{2006A&A...455L..17L,
2007MNRAS.376..809G}, the effect of which is about four times larger
than the tidal field \citep[e.g.][Figure 1]{2006A&A...455L..17L}.
For this reason, we focus on old stellar clusters and initialize our
simulated star clusters at an early epoch, before the galactic
environment would have formed. In order to simulate young clusters, a
galaxy simulation including baryons would be required. However, when
the old clusters formed, the GMC density in the star-forming
environment was likely very high, causing early disruption of low-mass
globular clusters \citep{2010ApJ...712L.184E}. This effect is not
included in our simulations.

In order to investigate a large number of clusters with a reasonable
amount of stars, we used the \cite{1986Natur.324..446B} tree code {\tt
Bonsai} for most of our star cluster simulations. We compared the
results for two distinct star clusters to similar simulations with the
direct $N$-body code {\tt ph4}. The results for both runs are similar
for both codes, the direct code showing enhanced mass loss around the
time of core collapse. However, the tree code fails to accurately
describe the inner structure of the star cluster, and requires the
distance between stars to be softened.

Another limit of our environment is its resolution. The spatial
resolution of CosmoGrid is given by the softening length employed (175
parsec). Forces that occur on a scale similar to or smaller than this
softening length are not accurately taken into account.  Likewise, the
mass resolution of dark matter particles in CosmoGrid is
$1.28\times10^5$\,\Msun, about 4 times larger than the initial mass of
our simulated clusters. The effects of tidal forces caused by a small
dark matter object passing at close range to our clusters (such as a
subhalo) are therefore limited, creating a possible bias against the
effect of such structures. However, this resolution effect would be
more important for baryons than it is for dark matter.

The benefit of using a large-scale cosmological simulation however, is
that the formation of a dark-matter halo is followed. A simulation
that only models a collision between galaxies would not take the
earlier history and distribution of star clusters into account.  In
order to have the benefits of both a cosmological environment and high
resolution, one could use re-simulation, where galaxies are simulated
at high resolution within a lower-resolution environment.

Another limit imposed by our use of a pre-calculated simulation is its
limited number of snapshots. We lack continuous information about the
tidal tensor. In order to prevent sudden changes in the tidal field,
we interpolate the tidal tensor between snapshots. However, it remains
impossible to accurately track sudden changes in the tidal field on
timescales shorter than our time resolution, such as those occurring
during halo mergers. 
Since the orbital periods of our clusters around the halo centre are
larger than this time resolution, we do not expect this to have a
large influence on the evolution of the tidal field. However,
short-lasting passages of nearby objects may not be taken into account
accurately, and a method in which the tidal field is sampled at more
intervals remains preferable.

In a follow-up article (Rieder et al., in prep.), we will apply the
method described in this article to the evolution of star clusters in
the disk of a simulated Milky Way-type galaxy.  In this follow-up, we
will address several of the limitations discussed above, especially
the lack of baryons and as a result the orbits and origins of the
clusters.

\section*{Acknowledgements}

It is our pleasure to thank the anonymous referee for very helpful
suggestions and comments that greatly helped to improve the article.
Also, we are grateful to Arjen van Elteren, Derek Groen, Inti
Pelupessy, Jeroen B\'edorf, Mark Gieles and Nathan de Vries for
support, interesting discussions and useful suggestions. 

This work was supported by NWO (grants IsFast [\#643.000.803], VICI
[\#639.073.803], LGM [\#612.071.503] and AMUSE [\#614.061.608]), NCF
(grants [\#SH-095-08] and [\#SH-187-10]), NOVA and the LKBF in the
Netherlands, and by NSF grant AST-0708299 in the U.S.  
T.I. is financially supported by MEXT HPCI STRATEGIC PROGRAM and
MEXT/JSPS KAKENHI Grant Number 24740115.
We thank the DEISA Consortium (EU FP6 project RI-031513 and FP7
project RI-222919) for support within the DEISA Extreme Computing
Initiative (GBBP project).

The Cosmogrid simulations were partially carried out on Cray XT4 at
Center for Computational Astrophysics, CfCA, of National Astronomical
Observatory of Japan; Huygens at the Dutch National High Performance
Computing and e-Science Support Center, SARA (The Netherlands); HECToR
at the Edinburgh Parallel Computing Center (United Kingdom) and Louhi
at IT Center for Science in Espoo (Finland).

\newcommand{\aj}{AJ}
\newcommand{\actaa}{Acta Astron.}
\newcommand{\araa}{ARA\&A}
\newcommand{\apj}{ApJ}
\newcommand{\apjl}{ApJ}
\newcommand{\apjs}{ApJS}
\newcommand{\ao}{Appl.~Opt.}
\newcommand{\apss}{Ap\&SS}
\newcommand{\aap}{A\&A}
\newcommand{\aapr}{A\&A~Rev.}
\newcommand{\aaps}{A\&AS}
\newcommand{\azh}{AZh}
\newcommand{\baas}{BAAS}
\newcommand{\caa}{Chinese Astron. Astrophys.}
\newcommand{\cjaa}{Chinese J. Astron. Astrophys.}
\newcommand{\icarus}{Icarus}
\newcommand{\jcap}{J. Cosmology Astropart. Phys.}
\newcommand{\jrasc}{JRASC}
\newcommand{\memras}{MmRAS}
\newcommand{\mnras}{MNRAS}
\newcommand{\na}{New A}
\newcommand{\nar}{New A Rev.}
\newcommand{\pra}{Phys.~Rev.~A}
\newcommand{\prb}{Phys.~Rev.~B}
\newcommand{\prc}{Phys.~Rev.~C}
\newcommand{\prd}{Phys.~Rev.~D}
\newcommand{\pre}{Phys.~Rev.~E}
\newcommand{\prl}{Phys.~Rev.~Lett.}
\newcommand{\pasa}{PASA}
\newcommand{\pasp}{PASP}
\newcommand{\pasj}{PASJ}
\newcommand{\qjras}{QJRAS}
\newcommand{\rmxaa}{Rev. Mexicana Astron. Astrofis.}
\newcommand{\skytel}{S\&T}
\newcommand{\solphys}{Sol.~Phys.}
\newcommand{\sovast}{Soviet~Ast.}
\newcommand{\ssr}{Space~Sci.~Rev.}
\newcommand{\zap}{ZAp}
\newcommand{\nat}{Nature}
\newcommand{\iaucirc}{IAU~Circ.}
\newcommand{\aplett}{Astrophys.~Lett.}
\newcommand{\apspr}{Astrophys.~Space~Phys.~Res.}
\newcommand{\bain}{Bull.~Astron.~Inst.~Netherlands}
\newcommand{\fcp}{Fund.~Cosmic~Phys.}
\newcommand{\gca}{Geochim.~Cosmochim.~Acta}
\newcommand{\grl}{Geophys.~Res.~Lett.}
\newcommand{\jcp}{J.~Chem.~Phys.}
\newcommand{\jgr}{J.~Geophys.~Res.}
\newcommand{\jqsrt}{J.~Quant.~Spec.~Radiat.~Transf.}
\newcommand{\memsai}{Mem.~Soc.~Astron.~Italiana}
\newcommand{\nphysa}{Nucl.~Phys.~A}
\newcommand{\physrep}{Phys.~Rep.}
\newcommand{\physscr}{Phys.~Scr}
\newcommand{\planss}{Planet.~Space~Sci.}
\newcommand{\procspie}{Proc.~SPIE}

\bibliography{references}
\bibliographystyle{mn2e}

\end{document}